\documentclass[sigconf,edbt]{acmart-edbt2020}

\def\BibTeX{{\rm B\kern-.05em{\sc i\kern-.025em b}\kern-.08em
    T\kern-.1667em\lower.7ex\hbox{E}\kern-.125emX}}

\usepackage{booktabs} 
\usepackage{algorithm}  

\usepackage{algorithmicx,algpseudocode}

\setcopyright{rightsretained}

\acmDOI{}

\acmISBN{XXX-X-XXXXX-XXX-X}

\acmConference[EDBT 2020]{22nd International Conference on Extending Database Technology (EDBT)}{March 30-April 2, 2020}{Copenhagen, Denmark} 
\acmYear{2020}

\newcommand{\ie}{\textit{i}.\textit{e}., }
\newcommand{\etal}{\textit{et al}. }

\begin{document}
\title{Cracking In-Memory Database Index: A Case Study for Adaptive Radix Tree Index}


\author{Gang Wu}
\affiliation{%
  \institution{Northeastern University}
  \institution{State Key Laboratory for Novel Software Technology,Nanjing University}
  \streetaddress{195 chuangxin Rd}
  \city{Shenyang} 
  \country{China}
}
\email{wugang@mail.neu.edu.cn}


\author{Yidong Song}
\affiliation{%
  \institution{Northeastern University}
  \streetaddress{195 chuangxin Rd}
  \city{Shenyang} 
  \country{China}
}
\email{2488951516@qq.com}

\author{Guodong Zhao}
\affiliation{%
  \institution{Northeastern University}
  \streetaddress{195 chuangxin Rd}
  \city{Shenyang} 
  \country{China}
}
\email{1690253916@qq.com}

\author{Wei Sun}
\affiliation{%
  \institution{Baidu}
  \city{Beijing} 
  \country{China}
}
\email{sunwei@baidu.com}

\author{Donghong Han}
\affiliation{%
  \institution{Northeastern University}
  \streetaddress{195 chuangxin Rd}
  \city{Shenyang} 
  \country{China}
}
\email{handonghong@cse.neu.edu.cn}

\author{Baiyou Qiao}
\affiliation{%
  \institution{Northeastern University}
  \streetaddress{195 chuangxin Rd}
  \city{Shenyang} 
  \country{China}
}
\email{qiaobaiyou@mail.neu.edu.cn}

\author{Guoren Wang}
\affiliation{%
  \institution{Beijing Institute of Technology}
  \city{Beijing} 
  \country{China}
}
\email{wanggr@bit.edu.cn}

\author{Ye Yuan}
\affiliation{%
  \institution{Northeastern University}
  \streetaddress{195 chuangxin Rd}
  \city{Shenyang} 
  \country{China}
}
\email{yuanye@mail.neu.edu.cn}




\renewcommand{\shortauthors}{}

\begin{abstract}
Indexes provide a method to access data in databases quickly. 
It can improve the response speed of subsequent queries by building a complete index in advance. 
However, it also leads to a huge overhead of the continuous updating during creating the index. 
An in-memory database usually has a higher query processing performance than disk databases and is more suitable for real-time query processing.
Therefore, there is an urgent need to reduce the index creation and update cost for in-memory databases. 
Database cracking technology is currently recognized as an effective method to reduce the index initialization time. 
However, conventional cracking algorithms are focused on simple column data structure rather than those complex index structure for in-memory databases. 
In order to show the feasibility of in-memory database index cracking and promote to future more extensive research, this paper conducted a case study on the Adaptive Radix Tree (ART), a popular tree index structure of in-memory databases.
On the basis of carefully examining the ART index construction overhead, an algorithm using auxiliary data structures to crack the ART index is proposed.
This makes it possible to build up an ART index step by step with incessant queries, and hence avoids the poor instant availability of a complete index which is constructed once and for all, but is time consuming. 
Furthermore, updating a cracking ART index is considered as well.
Extensive experiments show that the average initialization time of the ART cracker index is reduced by 75\%, and the query response time gradually approaches the original ART algorithm with the coming queries.
\end{abstract}

%
%



\maketitle

\section{Introduction}\label{sec:introduction}
With the increasing requirements for real-time transaction and analysis processing, traditional disk-based data management techniques are no longer applicable to these scenarios. 
In order to meet the requirements for real-time performance of critical business, in-memory database (IMDB) came into being. 
More and more flexible and efficient in-memory data storage and access methods are used to increase system throughput and reduce response time.

As an important component of the database system, indexing technology has always been a research hotspot in the field of IMDB. 
A large number of index structures for memory data are proposed.
Compared with early work such as T tree \cite{LehmanC1986:T-tree}, CSB+ tree \cite{Rao2000:csb+-tree}, and CSS tree \cite{RaoR1999:css-tree}, state-of-the-art in-memory indexes like FAST \cite{KimCSSNKLBD2010:fast-index}, Masstree \cite{MaoKM2012:mass-tree}, BwTree \cite{LevandoskiLS2013a:BwTree}, PSL \cite{XieCJOW2017:PSL} and ART \cite{LeisKN2013:ART-index} achieve better performance by making good use of concurrent synchronization and new hardware technologies \cite{XieCCMZ2018:evaluation-imdb-index}.
The ART (Adaptive Radix Tree) is a representative one of these indexes which shows a competitive small memory footprint and overall performance especially for dense dataset by building a trie structure tree with adaptive variable length type internal nodes and supporting efficient SIMD processing \cite{XieCCMZ2018:evaluation-imdb-index}. 

Although indexes can improve query efficiency, for IMDB, there are more details to be considered, as it is inherently designed to pursue more rapid system response. 
Firstly, the time overhead of index construction cannot be ignored. 
Obviously, a one-time construction of the entire index may cause the system to become unavailable for a short period of time. 
Secondly, the effectiveness of an index is largely determined by the actual query workload. 
Indexes built in the absence of query workload information are likely to be time-consuming, space-intensive, and what's more may even be of no use. 
Thirdly, index reconstruction due to data updates can also increase system response time.

The database cracking technology \cite{IdreosKM2007:DB-cracking} provides a way to solve these problems. 
Instead of building a complete index in the preprocessing stage, it builds and refines the index along with the query processing. 
Thus, the cracker index also enables query workload aware, and hence avoids the case of ineffective indexing. 
At the same time, such dynamic index construction mechanism can better adapt to data updates. 
The philosophy behind the database cracking is to delay unavoidable changes as far as possible \cite{IdreosKM2007:updating-cracked}.
However, as traditional database cracking technology was originally designed towards simple array alike column data structures, it alone is uncompetitive compared to modern IMDB indexes. 
Recent work \cite{SchuhknechtJD2016:evaluation-cracking} shows that the data lookup speed of ART index is 3.6 times faster than the traditional database cracking method after 1M queries. 
Therefore, an ideal solution might be applying database cracking technology to modern IMDB index construction to further improve the overall system responsiveness (considering the overhead of index construction and update, and the index effectiveness). 
To the best of our knowledge, there is currently no specific research on database cracking for complex index structures. 
Thus, it is necessary to study the feasibility and relevant general techniques of the complex index cracking.
For this reason, a preliminary case study is conducted on the ART index in this paper, which will be representative for in-memory index.

We studied the construction overhead of ART index and proposed an algorithm that cracks ART with the help of auxiliary data structures. The main contributions of this paper are as follows.
\begin{enumerate}
    \item We investigated the impact of data ordering on the construction of ART index and the range lookup performance of ART index. 
    We found that cracking technology can improve the construction of ART because ordered data come into being gradually during the cracking process; 
    \item We  proposed a cracking algorithm for the Adaptive Radix Tree with the help of auxiliary data structures which has a low index initialization overhead and guarantees to eventually form a complete ART index in the process of constant queries. 
    The algorithm is easy to implement and is applicable to other complex index structures that support range query. 
    Furthermore, in order to improve the update performance, caching and shuffling techniques are introduced. 
    \item Extensive experiments were conducted under different workloads on two datasets, \ie a synthetic dataset and the YCSB \cite{CooperSTRS2010:YCSB} benchmark. 
    The experimental results were analyzed to show the feasibility of the proposed cracking algorithm from the convergence speed, response time, and selectivity.
\end{enumerate}

The organization of this paper is as follows: 
Section 2 introduces related work; 
Section 3 discusses the feasibility of cracking ART index by analyzing the construction process of the index and the characteristics of query processing on it; 
Section 4 describes the design and implementation of the ART cracker in detail;
Section 5 shows and analyzes the experimental results; 
Finally, conclusions are given in Section 6.

\section{Related Work}\label{sec:relate-work}
More and more index structures are proposed for in-memory database systems.
Rao \etal \cite{Rao2000:csb+-tree} proposed the Cache-Sensitive B+-Tree (CSB+-Tree) that retains the good cache behavior of CSS-Trees while at the same time being able to support incremental updates.
Hankins \etal \cite{Hankins2003:csb-tree} explored the effect of node size on the performance of CSB+-trees and found that using node sizes larger than acache line size (i.e., larger than 512 bytes) produces better searchperformance. While trees with nodes that are of the same size as a cache line have the minimum number of cache misses, they found that TLB misses are much higher than on trees with large node sizes, thus favoring large node sizes.
Kim \etal \cite{KimCSSNKLBD2010:fast-index} proposed the Fast Architecture Sensitive Tree (FAST),a binary tree logically organized to optimize for architecture features like page size, cache
line size, and SIMD width of the underlying hardware.Additionally, they proposed to interleave the stages of multiple queries in order to increase the throughput of their search algorithm.
However, both  \cite{LeisKN2013:ART-index} and \cite{FelixMartin2013:uncracked} indicate that ART outperforms other main-memory optimised search trees such as CSB+-Tree  and FAST.

\begin{figure}[htp!]
   \centering
   \includegraphics[scale=0.35]{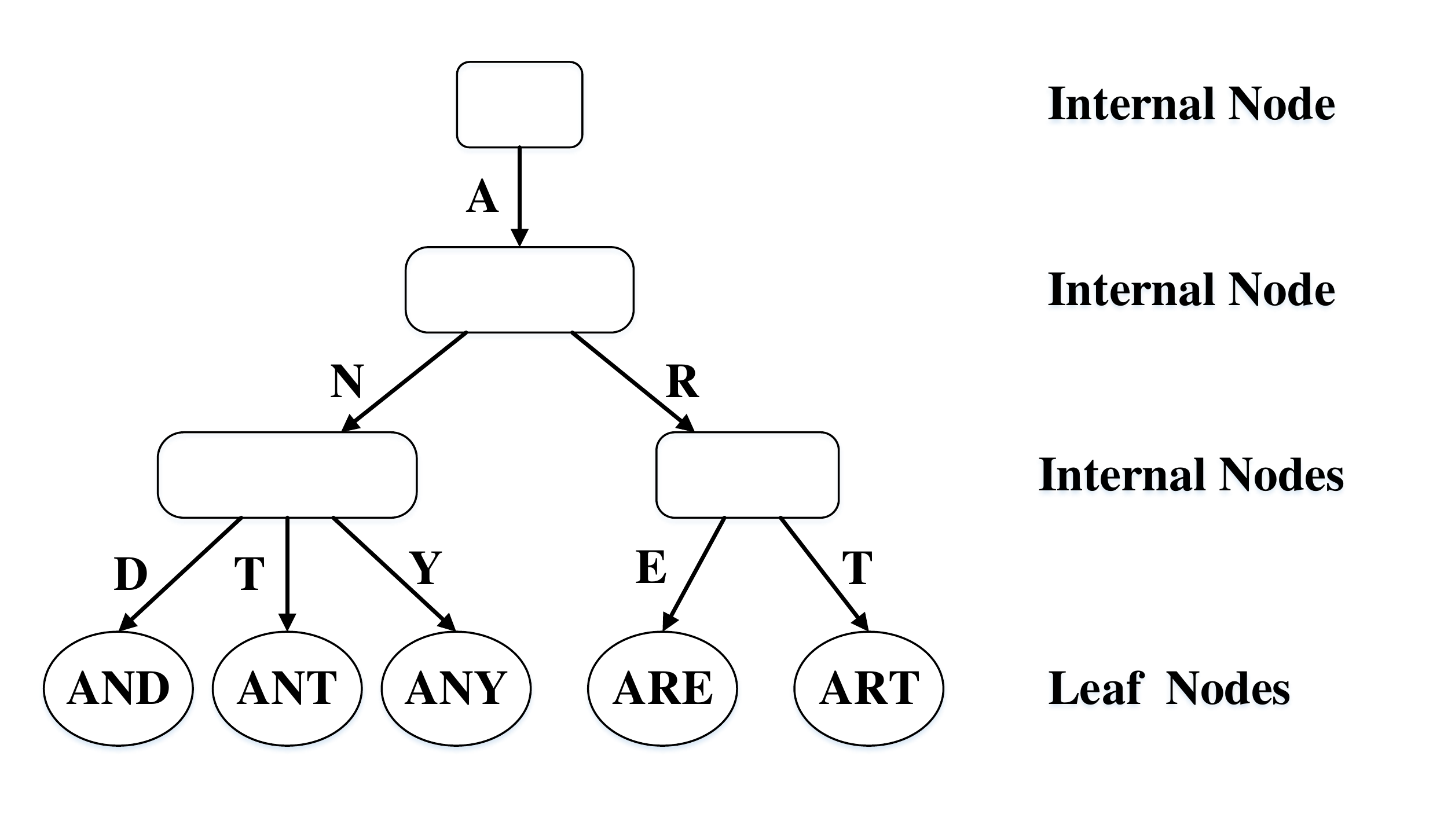}
   \caption{An example ART index structure \cite{LeisKN2013:ART-index}.}
   \label{fig:ART}
\end{figure}

Adaptive Radix Tree (ART) index was first proposed in \cite{LeisKN2013:ART-index}. 
As the example shown in Figure \ref{fig:ART}, ART index has two types of nodes where the internal nodes provide mappings from partial key to other nodes, and the leaf node stores the value corresponding to the keyword. 
The height of the tree is only determined by the maximum length of the indexed keyword rather than the number of indexed keywords, and all keys are sorted lexicographically. 
The capacity of an internal node changes adaptively with the inserted keywords, which makes ART cache aware, and query efficient. 
In addition, by employing the path compression and lazy expansion techniques, ART further reduces the space consumption. 
ART index has been integrated in an in-memory database system, HyPer \cite{KemperN2011:Hyper}, which shows better query performance than other in-memory database index struc-tures. 
Recently, Leis V \etal proposed the Optimistic Lock Coupling and Read-Optimized Write EXclusion (ROWEX) protocols to deal with the synchronization problem of ART index \cite{LeisSKN2016:ART-sync}, which further improve the performance of ART index and expand its application scope.

\begin{figure*}[htp!]
   \includegraphics[scale= 0.75]{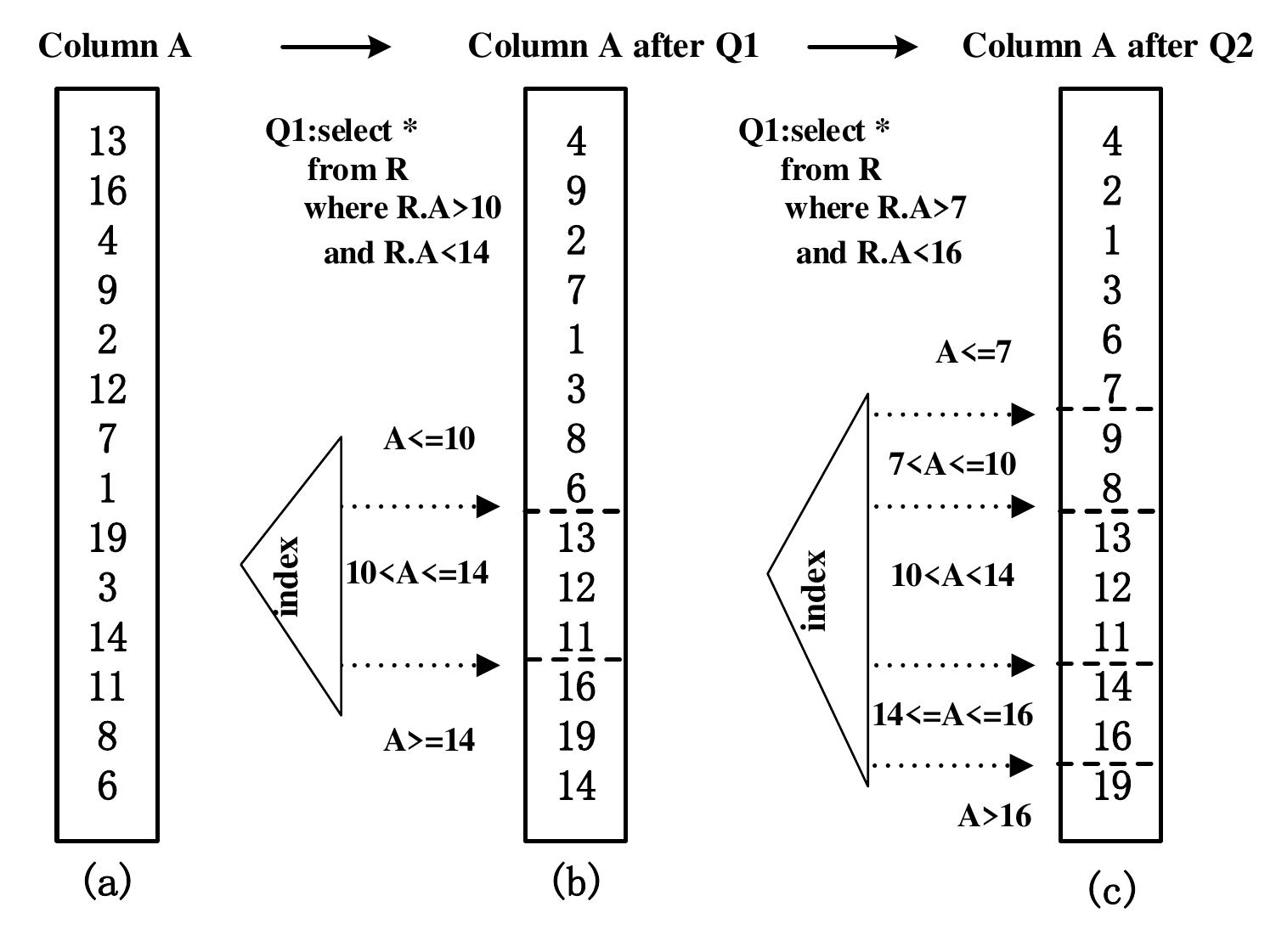}
   \caption{Database Cracking when executing two queries \cite{IdreosKM2007:DB-cracking}.}
   \label{fig:cracking}
\end{figure*}

The idea of database cracking was first proposed in \cite{KerstenM2005:cracking-DS} and simulated with SQL statements, which showed the application prospect of the algorithm. 
For an attribute to be cracked, the database cracking technology first makes a copy (called the cracker column) of the corresponding column, and then partially records the tuples in the cracker column into tuple clusters (called the column slices) according to the continuously arrived range queries on the attribute until the column is completely sorted. 
Figure \ref{fig:cracking} depicts the standard Database Cracking when executing two queries.
The tuples are clustered in three pieces from the range predicate of Q1.The result of Q1 is then retrieved as a view on Piece 2(\ie indexing 10 < A < 14).
Later, query Q2 requires a refinement of Pieces 1 and 3(\ie respectively indexing A > 7 and A < 16) and splitting each in two new pieces.More database cracking implementation details are discussed in \cite{IdreosKM2007:DB-cracking}. 

It shows good performance when the cracking algorithm is embedded in MonetDB \cite{IdreosGNMMK2012:MonetDB}. 
Recently, research on database cracking technology has become more extensive and in-depth.
In \cite{IdreosKM2007:updating-cracked}, Idreos \etal discussed the problem of database cracking update. 
In \cite{Halim:stochastic},Halim  \etal proposed the Stochastic Database Cracking,it alleviates the sensitivity of the cracking process to the kind of queries by introducing random physical reorganization steps for efficient incremental index-building, while also taking the actual queries into account.
In \cite{IdreosMKG2011:merging-cracking},Idreos  \etal proposes a series of Hybrid Cracking algorithms based on the combination of Database Cracking and Adaptive Merging \cite{Graefe2010:self-tuning,Graefe2010:adaptive}.
These hybrids are intended to meet both the database cracking design goal of minimizing initial per query overhead and also the adaptive merging design goal of exploiting the concept of runs and merges to converge quickly.
In \cite{IdreosKM2009:self-organizing-cs}, a number of technologies such as Sideways Cracking, Multi-projection Queries, and Tuple Reconstruction were proposed to solve the problem of cross-column query in column databases. 
The concept of tuple reorganization is put forward in this paper as well. 
Moreover, the authors summarized existing database cracking algorithms comprehensively in \cite{SchuhknechtJD2016:evaluation-cracking}, and further improved the original algorithm in terms of the convergence speed, robustness and parallelism, and suggested that the future development direction of database cracking is to adapt to the changes of in-memory index structures.

\section{Feasibility Analysis of ART Index Cracking} \label{sec:feasibility}
In this section, we will present the necessity and possibility of applying cracking technologies on ART index.

\subsection{ART Index Construction Overhead} \label{subsec:art-overhead}
During the construction of the index, ART recursively maps each part of the key to a child internal node along a path of the tree from the root. 
In order to efficiently manage the size of the internal node, four internal node types with different capacities are provided. 
An appropriate internal node type is selected according to the number of its children, \ie its fanout.

As we know, ART is a trie structure tree. 
It means that the tree structure remains the same regardless of the node insertion order. 
In other words, there will be no rebalancing of the tree. 
Although different key insertion orders lead to the same ART structure, we found that the order actually has a significant impact on the time overhead of index construction.

\begin{figure}[h]
   \includegraphics[width=\linewidth]{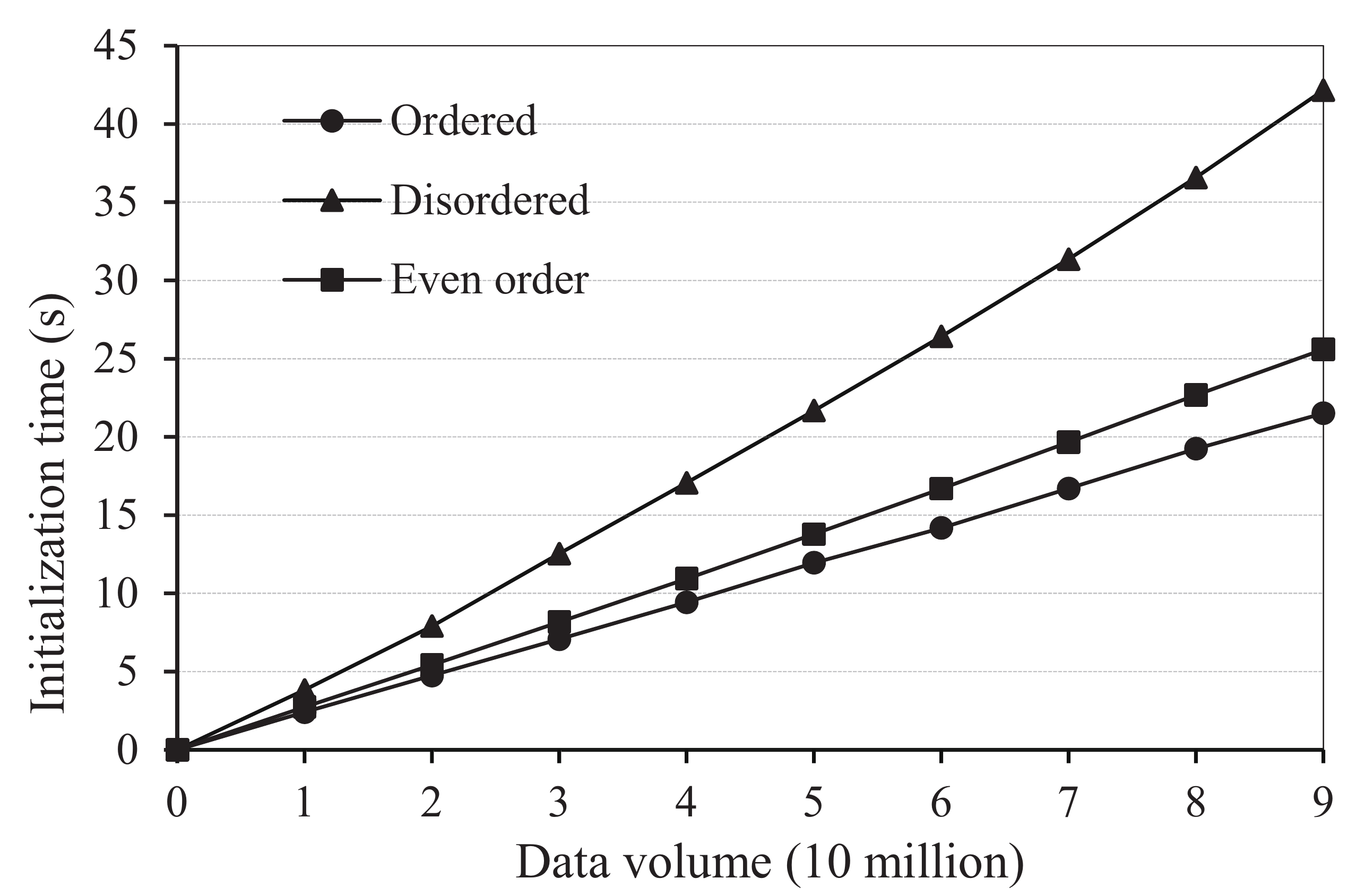}
   \caption{The impact of key insertion order on ART.}
   \label{fig:InsertOrderImapct}
\end{figure}

Figure \ref{fig:InsertOrderImapct} shows our experimental results of the impact of key insertion order on ART index construction. 
The experiment was conducted on a commodity server with an Intel(R) Xeon(R) CPU E7-4820 v4 @ 2.00GHz, 25600K L3 Cache, and 1TB RAM. 
We prepared three sequences of integer data. 
They have the same length $N>0$, but the elements are distributed in different orders. 
For the \emph{Ordered} sequence, all elements are arranged in ascending order within the range [1, $N$]; for the \emph{Disordered} sequence, all elements are randomly arranged within the range [1, $N$]; and for the \emph{Even order} sequence, all even elements are arranged in ascending order within the range [1, 2$N$]. 
The index construction time increases approximately linearly with the amount of data inserted for all three cases. For the same amount of data, the comparison of the time overhead of the three is Disordered > Even order > Ordered. 
And their gaps also increase with the amount of data. 
The index construction time for Disordered even becomes twice of that for the Ordered when data amount to reaching 90 million. 
Moreover, the experiment also shows that larger data intervals (the case of Even order here) require more index construction time.

The experiment is a good illustration of the fact that the construction of the ART index is sensitive to the key insert order. 
As we know, the data exchange between main memory and CPU cache is through cache lines, \ie fixed size (usually 32 or 64 bytes) of blocks. 
Successively inserted disordered keys or evenly ordered keys may cause them to span more ART tree nodes, which make more different nodes to be visited in an insertion, and hence cache misses occur more frequently. 
Since accessing the cache is much faster than accessing the main memory, the cases of disordered insertion and large interval insertion inevitably need more index construction time. 

Since keys are sorted gradually according to the continuously arrived queries during the process of cracking, it is worthwhile applying the cracking technology to the construction of ART index.

\subsection{ART Range Query} \label{subsec:art-range}
The process of database cracking is driven by continuously issued range queries.
Therefore, it should be ensured that any data structure to be cracked will support range queries. 
According to \cite{LeisKN2013:ART-index}, the ART index supports not only point queries but also range scans. 
As child pointers are sorted in an internal node, the range scan can be performed efficiently by returning all leaf nodes in a subtree between lower and upper bounds of a range.

For conventional database cracking technology, the overall query performance is bounded by the computational complexity of the binary search on an ordered array, \ie O(log$N$) where $N$ is the amount of data in the array. 
Fortunately, the number of comparisons is only related to the path length of the key in the ART index tree rather than the number of keys indexed, and the path length is usually much smaller than log$N$. 
From this point of view, to index a column, it seems faster to start with cracking an ART index on the column than to crack the column directly.

\section{ART Index Cracking} \label{sec:art-cracking}
In this section, we propose an ART index cracking method which is based on the assumption that the time overhead of index construction and maintenance cannot be ignored for large scale IMDB where real-time query response is critical in the meanwhile. 
Although a completed ART index has excellent query response performance, the initialization phase to construct the index from scratch makes it unavailable for the scenarios that require instant query response at any time. 
The ART index cracking algorithm contributes a possible solution to the performance trade-off between instant query and overall query. 
It can also ensure the index effectiveness under unknown query workload. 
The basic idea behind the ART index cracking algorithm is to use auxiliary data structures to gradually construct an ART index rather than a sorted column along with the range queries issued on the column. 

\subsection{Components for ART Index Cracking} \label{subsec:components}
\begin{figure}[htp!]
   \includegraphics[width=\linewidth]{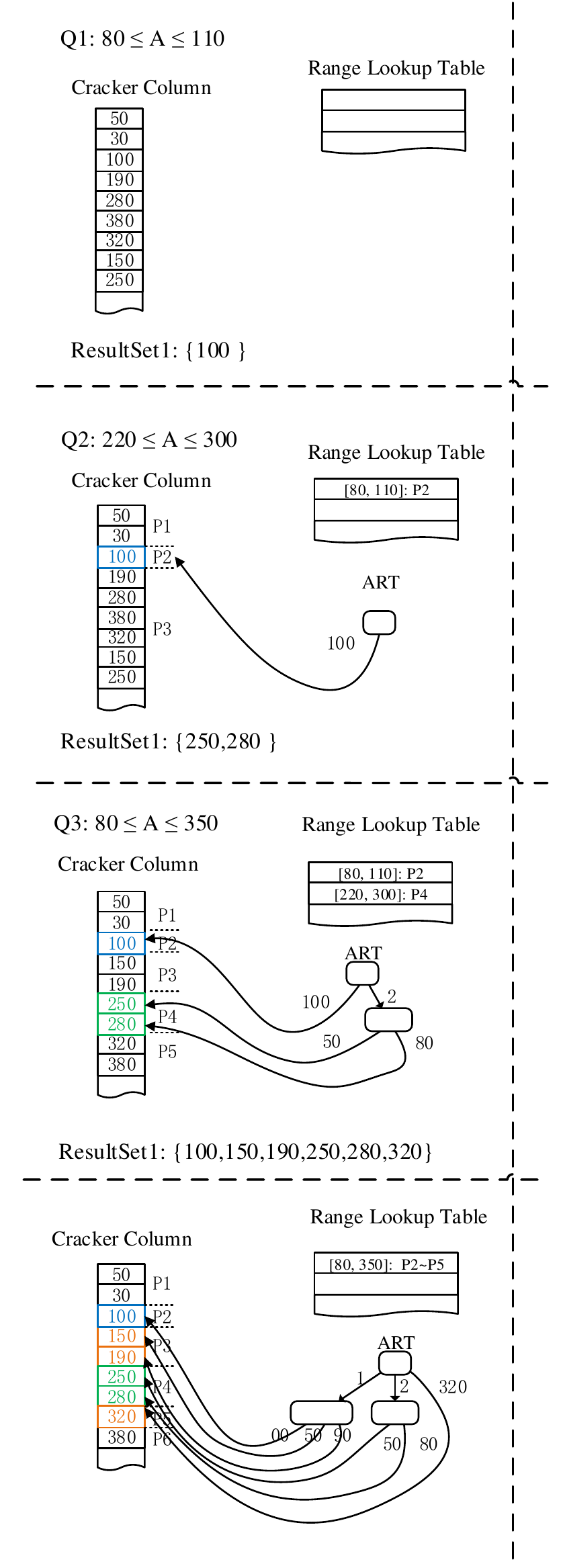}
   \caption{The process of ART index cracking with three range queries.}
   \label{fig:RangeQuery2}
\end{figure}

According to \cite{IdreosKM2007:DB-cracking}, \emph{Cracker Column} and \emph{Cracker Index} are the basic components of the conventional database cracking algorithm. 
Since our aim is to obtain a completed ART index rather than a sorted column, we assemble a cracker index with an ART index and an auxiliary data structure for maintaining the information of column data organization.
The components of our ART index cracking algorithm are illustrated in Figure \ref{fig:RangeQuery2}.

\begin{itemize}
\item[-] \emph{Cracker Column} is a copy of the database column from which an ART index is constructed. 
From the perspective of the ART index, it holds an array of keys to be indexed.
\item[-] \emph{Cracker Index} is an auxiliary index structure to locate and organize data in the cracker column into column slices which are the clusters of data on the column identified by range intervals, and to sort the column slices in a sorted order. 
In conventional implementation, any structure that supports search in sorted order (at least a total preorder) is available, such as the red-black tree and the AVL tree.
As analyzed in \ref{subsec:art-range}, ART is available this case.
In order to provide a sound ART cracking function, the cracker index here consists of two parts.
    \begin{itemize}
        \item[-] \emph{ART Index} is what to be gradually constructed from scratch according to the continuously arrived range queries on the column. 
        Internal nodes collapsing techniques are welcome. 
        \item[-] \emph{Range Lookup Table} is used to record those column slices which have already been indexed in the ART so that the cracking algorithm can decide whether to return the query results directly through the ART index or to perform the standard cracking operation through the cracker index. 
        To achieve this goal, merged historical query ranges and the corresponding covered column slices are recorded as the keys and values in the range lookup table. 
    \end{itemize}
\end{itemize}

The following points should be noted. 
Firstly, the column slices are dynamic during the entire cracking process. 
A range query on the cracker column may cause a column slice to be created by splitting or removed by merging existing slices. 
Secondly, any column slices covered by previous range queries has already been kept in the range lookup table. 
This ensures the improvement of overall query performance based on the strategy that uses the ART index for related query evaluation as much as possible.

\subsection{ART Index Cracking with Range Queries} \label{subsec:cracking}
Figure 3 illustrates the case of ART index cracking with three consecutive range query processing.
For the convenience of discussion, suppose queries occur on column $A$ of table $R$, and the data type of values in column A is Integer.

Since the range lookup table is empty when the first range query Q1 (80 $\leq$ A $\leq$ 110) is issued, a cracker column will be created by making a copy of the original column A. 
Then an ART index is constructed to index all data in the cracker column according to the lower bound and the upper bound of the range query, \ie 80 and 110 respectively. 
Hence, the cracker column is split and rearranged into three column slices, P1 (A < 80), P2 (80 $\leq$ A $\leq$ 110), and P3 (A > 110), which guarantees a sorted order in the cracker column between the column slices, \ie P1 $<$ P2 $<$ P3, \ie the lower bound index of P2 larger than the upper bound index of P1 and the upper bound index of P2 lower than the lower bound index of P3. 
All data in P2 satisfy the query, hence are returned as the result set.
Finally, the queried range [80, 110] and the covered column slice P2 are recorded in the range lookup table as a key-value pair.

Since there is no intersection between the query ranges of Q2 (220 $\leq$ A $\leq$ 300) and any queried ranges recorded in the range lookup table ([80,110] currently), a similar cracking process was conducted as that of Q1. 
Firstly, it can be infer that the result set must be included in the column slices larger than P2, \ie P3 according to the key-value pair recorded in the range lookup table, because the lower bound of Q2 is larger than the upper bound of the only key, \ie $110 < 220$. 
This results in two more column slices separated from P3, \ie P4 (220 $\leq$ A $\leq$ 300) and P5 (A > 300).
The result set consisting of all data from P4 is returned and further inserted into the ART index. 
The range lookup table is then updated with [220, 300] as a key and the corresponding covered column slice P4 as its value. 

For the range query Q3 (80 $\leq$ A $\leq$ 350) whose range stretches over P2, P4, P3 and part of P5, with the former two column slices having been recorded in the range lookup table, while P3 and partial P5 not been visited and sorted yet, the algorithm will form a union result set comprised of both the search results that are obtained directly from the ART and the search results that are obtained from P3 and P5 on cracker column. 
Note that a new column slice P6 will be split from P5 with respect to the upper bound value 350. 
At the end of the query process of Q3, those data in P3 and new P5 are sorted and inserted into the ART index, and the ranges in the range lookup table are replaced by a new key-value pair, \ie an interval [80, 350] and a merged column slice across P2, P3, P4, and P5, because both [80,110] and [220, 300] can be covered by [80, 350].

In this way, the complete ART index is gradually constructed by continuously arrived range queries. 
Through the analysis of the above algorithm execution process, it can be seen that the selectivity of range queries is an important factor to the performance.
On the one hand, the fewer tuples each query selects, the faster the query response will be, but the longer it will take to construct the complete ART index. 
On the other hand, the more tuples each query selects, the slower the query response will be, but the shorter time it will cost to construct the complete ART index.
Although it seems impractical that the build speed of ART index depends on the range query selectivity, we should be aware that it coincides with the philosophy of database cracking, \ie delaying unavoidable changes as far as possible.
It means that the ART index creation of cracking ART reflects the practical workload.
The analysis of the impact of selectivity will be discussed in detail in the experimental section.

\begin{algorithm}[htbp]
	\caption{ART Index Cracking with Range Queries}
	\label{alg:ARTCracking}

	\begin{algorithmic}[1]
	\Require \qquad CrackerColumn $col$, RangeLookupTable $tbl$, \qquad \qquad \qquad ARTIndex $art\_idx$
	\Function {art$\_$cracking}{[$lb$, $ub$]}
	    \If{$tbl$.empty()} 
	        \State $art\_idx$.init() 
	        \State $col$.init()
	    \EndIf
	    \State $hit\_range \gets \emptyset, hit\_slices \gets \emptyset, rs\_art \gets \emptyset$
	    \ForAll {$entry \in tbl$}
	        \State $hit\_range \gets entry.key \cap [lb, ub]$
	        \If {$hit\_range \not= \emptyset$ }
	            \State $rs \gets rs \cup art\_idx.$scan($hit\_range$)
	            \State $hit\_slices \gets hit\_slices \cup entry.values$
	        \EndIf
	    \EndFor
	    \State $new\_slices \gets \emptyset$
	    \ForAll {$slice \in (col.\text{slices()} - hit\_slices)$}
	        \If{$slice \cap [lb, ub] \not= \emptyset$}
    	        \State $new\_slice = slice \cap [lb, ub]$
    	        \State $new\_slices \gets new\_slices \cup \{new\_slice\}$
    	        \State $rs\_db \gets$ \Call{db$\_$cracking}{$new\_slice$}
    	        \State $art\_idx$.insert($rs\_db$)
    	        \State $rs \gets rs \cup rs\_db$
	        \EndIf
        \EndFor
        \State $key \gets [lb,ub]$, $value \gets new\_slices$
        \State $tbl.\text{insert\_and\_merge}(key, value)$
        \State \textbf{return} $rs$
    \EndFunction
	    
	    
	\end{algorithmic}
\end{algorithm}

As shown in Algorithm \ref{alg:ARTCracking}, given a range query $[lb, ub]$ on a column as the input, the algorithm outputs a set of values $rs$ as the range query result on the column $col$ while cracking the ART index $art\_idx$.
The algorithm of ART index cracking consists of four phases, \ie the initialization phase (line $2\sim 5$), the range search in the ART index (line $6\sim 13$), 
the cracking phase (line $14\sim 23$), and the finishing phase (line $24\sim 26$).

If the the range lookup table $tbl$ is empty, \ie there is not any range recorded in the table, the algorithm will first initialize the ART index $art\_idx$ and the cracker column $col$ (line 3 and 4). 

After the inialization, the algorithm performs a range search in the ART index guided by the range lookup table.
It first finds out all entries recorded in $tbl$ whose key has non-empty intersection $hit\_range$ with $[lb, ub]$ (line 8).
For each found entry, perform the range search $hit\_range$ in the ART index and make $rs$ union with current results (line 10).
Corresponding column slices in each found entry are collected in a set $hit\_slices$ for use in the next phase (line 11).

Entering the cracking phase, the algorithm iterates through all the column slices in the cracker column $col$ that are not indexed in $art\_idx$ but intersected with $[lb, ub]$ (line 15 and 16).
Note that there is only one complete slice across the entire cracker column initially.
All such intersections are collected in a set $new\_slices$ for updating the range lookup table $tbl$ in the last phase (line 17 and 18).
The traditional database cracking function \textbf{DB\_CRACKING()} is called for cracking each such intersection (line 19).
After calling \textbf{DB\_CRACKING()}, new column slices are sperated from the original ones, all values in the intersection are sorted and returned in $rs\_db$.
Furthermore, values in $rs\_db$ are inserted into the ART index $art\_idx$ (line 20), and union is made between $rs$ and $rs\_db$ (line 21).

In the finishing phase, a key-value pair is constructed by taking $[lb, ub]$ as key and all new column slices $new\_slices$ as value (line 24).
The key-value pair is inserted into the range lookup table $tbl$ and merged with the existing entries (line 25).
Taking the entry in the lookup table in Figure \ref{fig:RangeQuery2} after executing Q3 for example, the previous two entries with keys [80, 110] and [220, 300] are merged with the new key-value pair [80, 350]:\{P3, P5\} to be a new and only one entry [80, 350]:\{P2, P3, P4, P5\}.
Finally, $rs$ is returned as the result set. 

\subsection{ART Index Cracking with Updates} \label{subsec:updats}
The algorithm of ART index cracking with range query simply considers the case where the column to be queried and indexed is read-only. However, updates (insertions and deletions) to the columns are inevitable in the actual database application scenario. Index update overhead brought by this may result in a decrease in database system response performance. Preliminary idea to this problem is discussed below.

\subsubsection{Caching}\label{subsubsec:caching}
The basic idea is to delay the update operations by caching them until the arrived range query has an intersection with any cached update operations or until the cache is full. An exception is that the updated data fall within any range in the range lookup table, which will be handled immediately by ART index. In this way, write lock contention is effectively reduced. 

For this purpose, a sorted list is used to cache the updated data and their corresponding operation type (either insertion or deletion) according to the data value. 
Assume that this simple cache structure is rather small compared with the entire column, so the cost of sorting is negligible. 
For the same data value, a deletion after an insertion will both be removed from the list, while an insertion after a deletion will cover the previous deletion. 
Once the arrived range query has an intersection with any cached update operations, all data in the intersection will be inserted or deleted from the ART index, while all data in the query range which are not in the sorted list cache will be cracked as stated in Section \ref{subsec:cracking}. 
Finally, those update operations that have been performed in the ART index are removed from the sorted list.

For the case when the sorted list (cache) is full, a range query between the lower and upper bound of the data in the sorted list is constructed and issued actively by the algorithm to trigger an update process on the ART index. 
The process after that is just the same as the previous case.

It should be noted that the data to be deleted in the sorted list cache should be carefully examined and not to be returned in the result set of a range query.

Although there is a more straightforward approach to perform updates directly on the cracker column, the overhead of updating the cracker column includes not only the maintenance of the cracker column but also the maintenance of the ART index and the range lookup table.

\begin{algorithm}[htbp]
	\caption{ART Index Cracking with Insert}
	\label{alg:ARTCracking-insert}

	\begin{algorithmic}[1]
	\Require \qquad CrackerColumn $col$, RangeLookupTable $tbl$, \qquad \qquad \qquad ARTIndex $art\_idx$, Cache $cache$ 
	\Function {insert$\_$art$\_$cracking}{$lb$, $ub$,$value$}
	    \State $next$ = point at the last position of $col$  	
	    \State $cur$ = point at the first position of last $slice$
	    \If{$value \in tbl$} 
	        \State $art\_idx$.insert() 
	    \Else 
	        \State $cache$.insert()
	    \EndIf
	    \If{$cache \text{ is full}$}
	        \State $result \gets cache$
	    \Else
	        \State $result \gets cache \cap [lb, ub]$
	    \EndIf
	     \If{$result \not= \emptyset$ }
	        \ForAll{$entry \in  result$}
	            \While{$entry \notin cur\_slice(cur)$}  
	                \State $col[next] \gets col[cur]$
	                \State $next \gets cur$
	                \State $cur \gets \text{point at previous slice of $cur$}$
	            \EndWhile
	            \State $col.\text{insert}(entry,cur)$
	            \State $tbl.\text{update}()$
	            \State $cache.\text{update}()$
	            \State $art\_idx.\text{updare}()$
	        \EndFor
	    \EndIf
    \EndFunction
	\end{algorithmic}
\end{algorithm}

\subsubsection{Shuffling}\label{subsubsec:shuffling}
Obviously, the maintenance overhead of the cracker column triggered by the update is not negligible.
In order to ensure the existing sorted order in the slices of the cracker column, a naive method is to move all the data behind the position to be inserted or deleted backward or forward respectively.
A more practical technology, \emph{shuffling}, has been introduced in \cite{IdreosKM2007:updating-cracked}.

\begin{figure}[htp!]
   \includegraphics[scale=1.3]{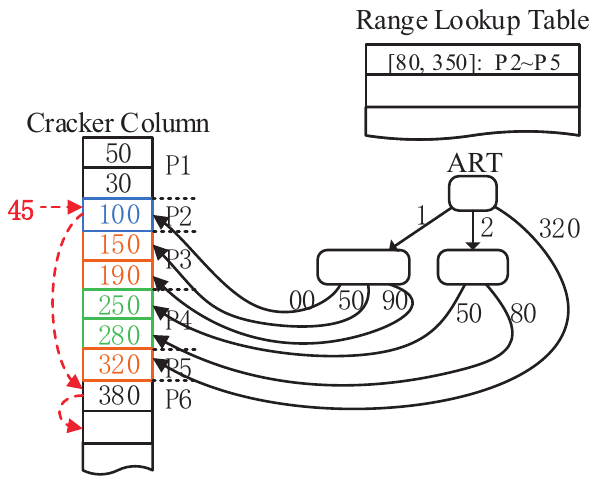}
   \caption{Shuffling of the cracker column after the insertion of 45.}
   \label{fig:Insert}
\end{figure}

Pseudo code for ART cracking with insert is shown in  Algorithm \ref{alg:ARTCracking-insert}. The algorithm of ART index cracking with insert consists of two phases, \ie the cache insert phase (line $4\sim 8$) and the shuffing a cracker column in range search (line $9\sim 25$), the necessary data is initialized in lines 2 and 3.

If the value to be inserted is in the range lookup table  $tbl$ , it can be inserted directly into the ART index  $art\_idx$ (line 5), otherwise it is inserted into the cache $cache$ to delay the update (line 7). 

The data in the cache $cache$ is  handled in (line $9\sim 13$).
If the cache is full, all data needs to be inserted(line 10), otherwise the intersection of the cache and the reached range query $[lb, ub]$ is processed(line 12). For all results $result$ that satisfy the condition,  move the element through shuffling a cracker cloumn(line $16\sim 20$ and get the position to insert.Then the value is inserted into the cracker cloumn and the range table $tbl$ (line 21), cache $cache$(line 23),the ART index $art\_idx$ (line 24) are updated.


The Algorithm \ref{alg:ARTCracking-insert} is explained below by an example as shown in Figure~\ref{fig:Insert}.When 45 in the cache is to be inserted, it first searches in the range lookup table.
Since 45 is smaller than the lower bound of the key ([80, 350]) of the only entry, it can be inferred that 45 should be inserted in the column slice P1.
Hence, according to the shuffling algorithm \cite{IdreosKM2007:updating-cracked}, for each lower bound value of the column slices that higher than P1, it is moved to the position of the lower bound of the next column slice.
As column slices P2, P3, P4, and P5 are all indexed in the ART index, and they are merged as one complete interval in the range lookup table, the shuffling process is actually only happens in P6 and P2.
Finally, 45 is inserted in the absent position that originally was owned by 100.
Although 100 is moved to the position behind 320 during shuffling, the sorted order of data in [80,350] is still maintained through the ART index.
In this way, the shuffling technology only needs two movements compared to the naive method requiring 7 movements.

Apparently, the shuffling technology can be adapted to deletion operations in a similar way.

\section{Experiments}\label{sec:experiments}
In this section, the advantages and disadvantages of standard cracking, standard ART, and ART cracking algorithms in various situations are compared through experiments.

The experimental settings are as follows. 
All algorithms were implemented in C++ language. 
All experiments were conducted on an Intel(R) Xeon(R) CPU E7-4820 v4 @ 2.00GHz server. 
The server has 512GB RAM and 25600K L3 Cache. 
The operating system is Ubuntu 18.04.4 LTS (GNU/Linux hp50 4.15.0-38-generic x86\_64).
All experiments were performed in memory to simulate the in-memory database.

In order to examine the performance of algorithms under different scenarios, we designed two sets of experiments.

The first set of experiments follow the experimental design and setting from \cite{IdreosMKG2011:merging-cracking} where the dataset is synthetic and the range query pattern is as follows. 
If there is no special statement, the experiment selects $N$ non-repetitive integers randomly distributed between $[1, N]$ where $N$ equals $10$M, the query selectivity $S$ is $0.0001N$, and the workload type is random.

\textbf{SELECT} $A$ \textbf{FROM} $R$

\textbf{WHERE} $A \geq low$ \textbf{AND} $A \leq high;$

The second set of experiments were conducted on YCSB (Yahoo! Cloud Serving Benchmark) which is a popular benchmark for evaluating different key-value and cloud database \cite{CooperSTRS2010:YCSB}.

\subsection{Space overhead}\label{subsec:space}
\begin{figure}[htp!]
   \includegraphics[width=\linewidth]{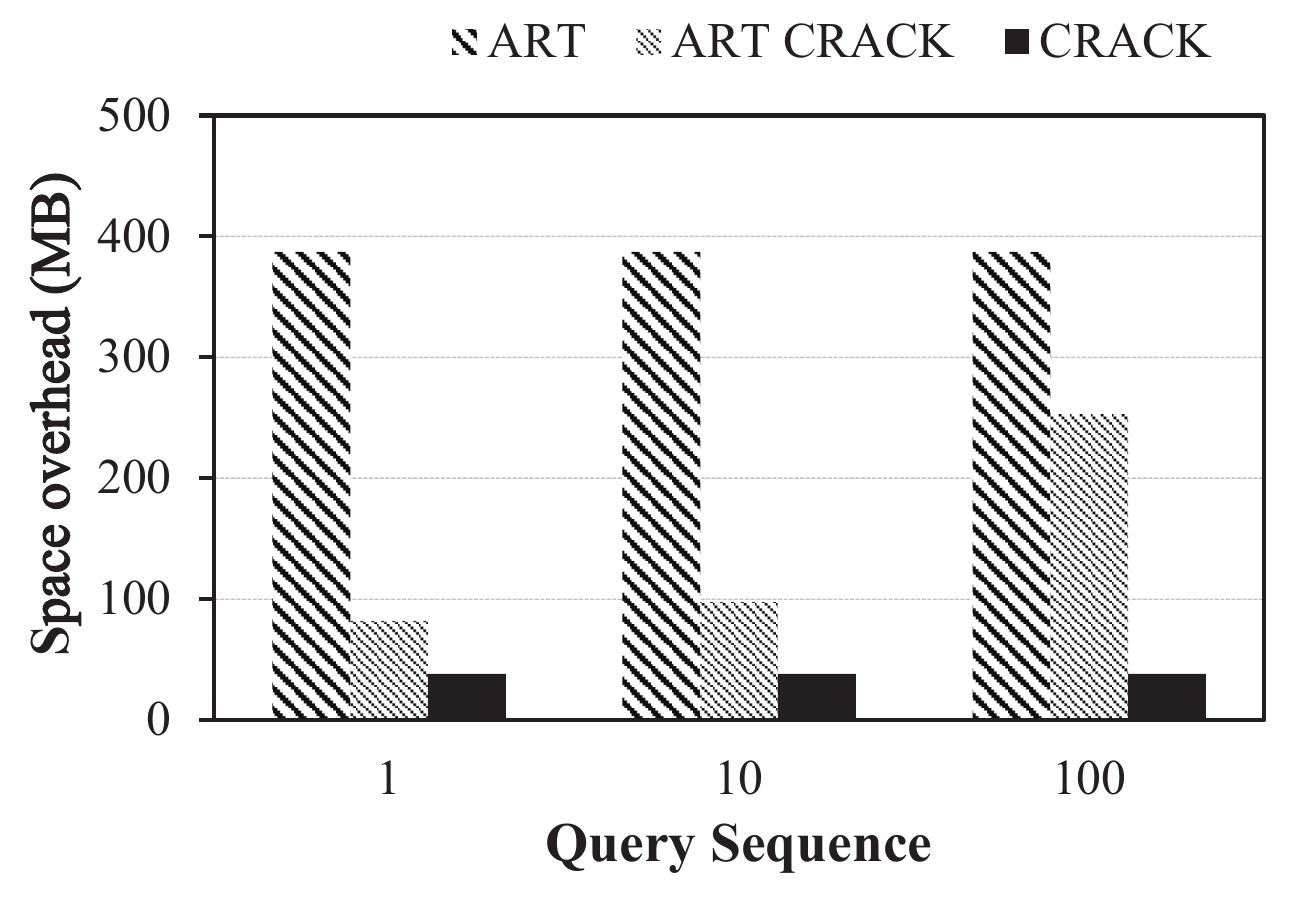}
   \caption{Space overhead comparisons when selectivity $S$= 0.005}
   \label{fig:space-fixed}
\end{figure}

\begin{figure}[htp!]
   \includegraphics[width=\linewidth]{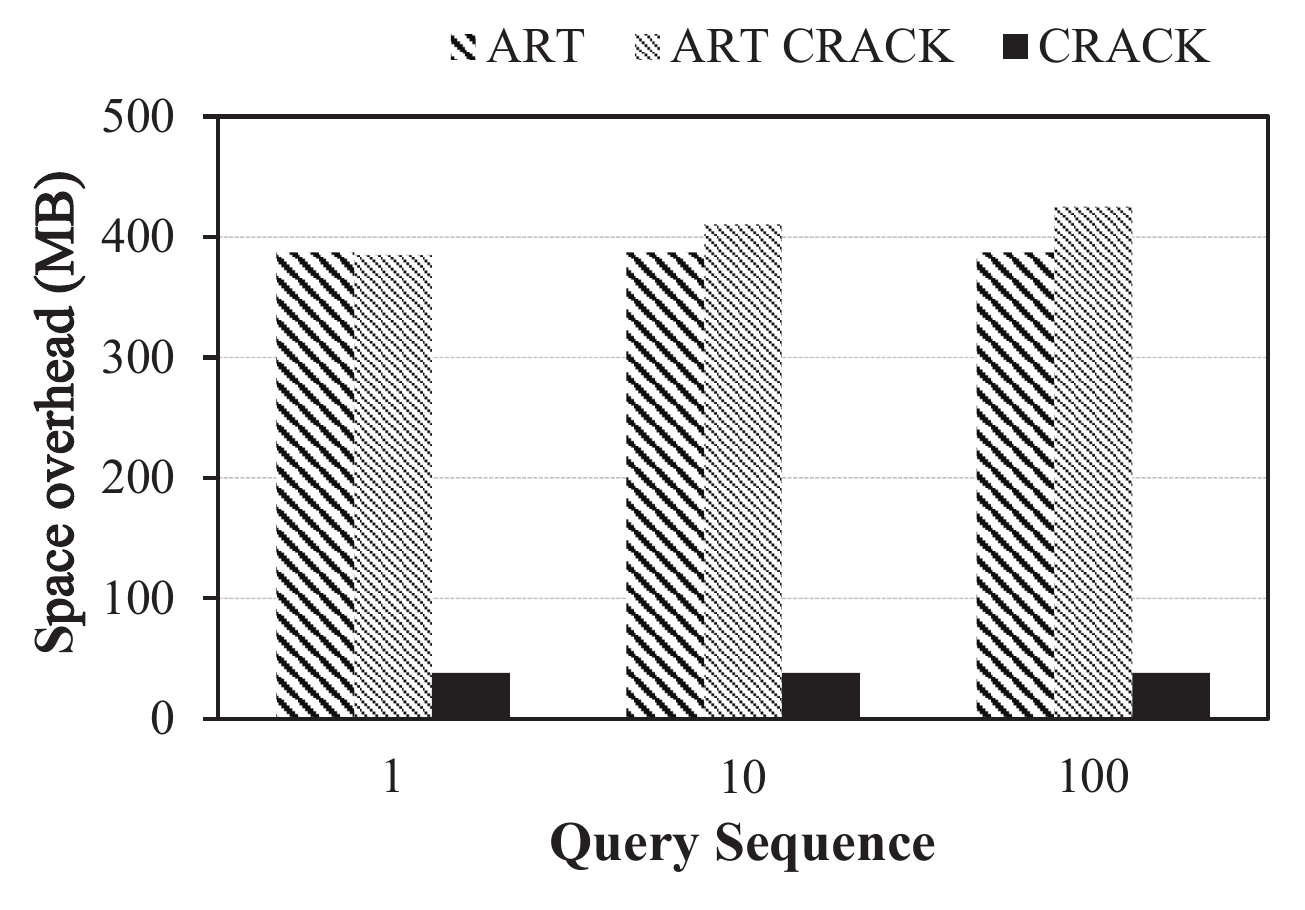}
   \caption{Space overhead comparisons for random query ranges}
   \label{fig:space-random}
\end{figure}

In order to compare the space overhead of the standard ART, the standard database cracking, and the ART cracking.
We did two experiments on the synthetic dataset with the default setting, \ie $N=10$M and non-repetitive integers randomly distributed between $[1, N]$.
For the first experiment, as shown in Figure~\ref{fig:space-fixed}, the lower bound and upper bound of each query range were selected with a fixed selectivity $S=0.005$.
For the second experiment, as shown in Figure~\ref{fig:space-random}, the lower bound and upper bound of each query range were generated randomly between $[1, N]$.

As demonstrated in both experiments, the standard ART and our ART cracking consume more space than the standard database cracking, because the index structure of ART is more complicated than that of database cracking (usually AVL or RBTree).
According to Figure~\ref{fig:space-fixed}, the space overhead of ART cracking increases with the number of queries and has a trend over that of ART.
Different from cracking methods, the space overhead of ART keeps unchanged.
Comparing Figure~\ref{fig:space-random} and Figure~\ref{fig:space-fixed}, we can also find that the space overhead grows faster in a random query range mode.
The reason is that there is a higher probability of generating a large query range for the random mode than that for the fixed selectivity mode.
As stated in Section~\ref{subsec:cracking}, a larger query range means more data to be indexed, and hence a larger space consumption. 

\subsection{Response Time}\label{subsec:response-time}
\begin{figure}[htp!]
   \includegraphics[width=\linewidth]{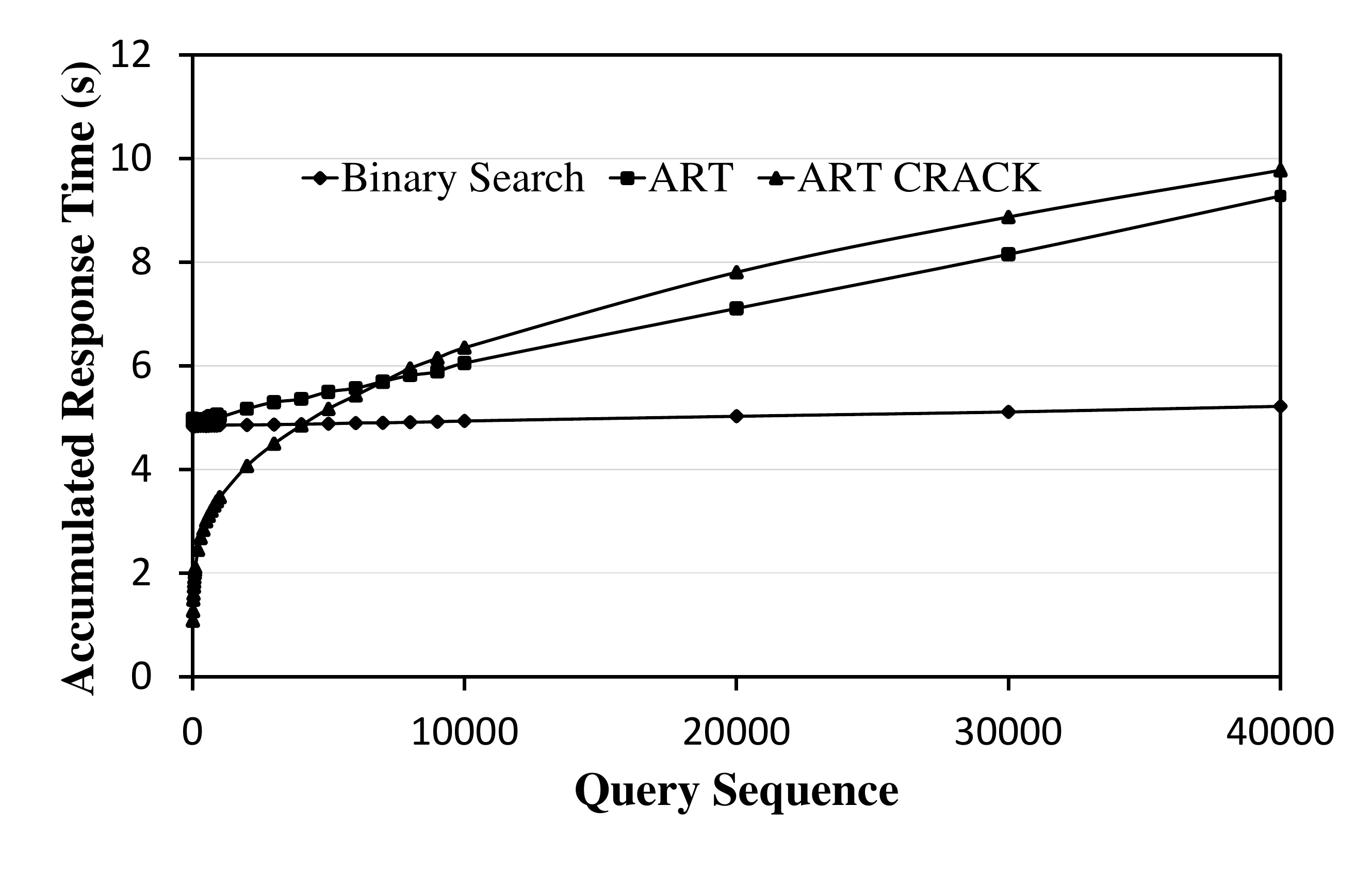}
   \caption{Response time comparisons for three algorithms}
   \label{fig:response-time}
\end{figure}

In Figure~\ref{fig:response-time}, we compare the response time of the standard ART, the ART cracking algorithm and the binary search algorithm in the synthetic dataset and random query mode.
The binary search algorithm applies the quick sort to sort the array first, and then use the binary search to find the result.
We have the following observations.

Firstly, compared with the binary search and the standard ART algorithm, the initialization time of the binary search algorithm and the standard ART algorithm is longer than that of the ART cracking algorithm.
This is due to the fact that the binary search algorithm needs to sort the data first, and the standard ART takes a lot of time to initialize the ART index as well. 
With the increase of the number of queries, the response time of standard ART and ART cracking algorithm increases linearly, while the increase of the response time of the binary algorithm is not significant.

Secondly, when the number of queries exceeds 5000, the response time of the ART cracking algorithm exceeds the binary search algorithm because of the good performance advantage of using binary search to perform range queries on ordered arrays.
As can be seen from the figure, when the number of queries exceeds 8000, the response time of the ART cracking algorithm exceeds the standard ART.
The reason is that ART cracking algorithm still needs a part of time to build a complete ART index, but the maximum difference between ART cracking algorithm and standard ART in response time is less than 1 second, and this difference is gradually reduced as ART is built up.

In summary, ART cracking algorithm can avoid building a complete ART index at one time and has a relatively low initialization cost.
At the same time, the hot data query is real-time, \ie when the query arrives, it avoids the waiting overhead due to the long time to initialize the ART.

\subsection{Selectivity}\label{subsec:selectivity}
The above experiments merely show the excellent performance of the ART cracking algorithm under a fixed selectivity. 
However, the impact of change to the selectivity cannot be ignored. 
On the one hand, when the selectivity is high, the algorithm converges quickly, while the initialization time is long, and the complete index structure is established promptly which fails to show the advantages of the ART cracking algorithm. 
On the other hand, when the degree of selection is low, the algorithm takes a long time to converge, while the initialization time is short. 
Especially for the case of querying hot data, though the query is increased, the cost of subsequent index maintenance is rather small.

\begin{figure}[htp!]
   \includegraphics[scale=0.31]{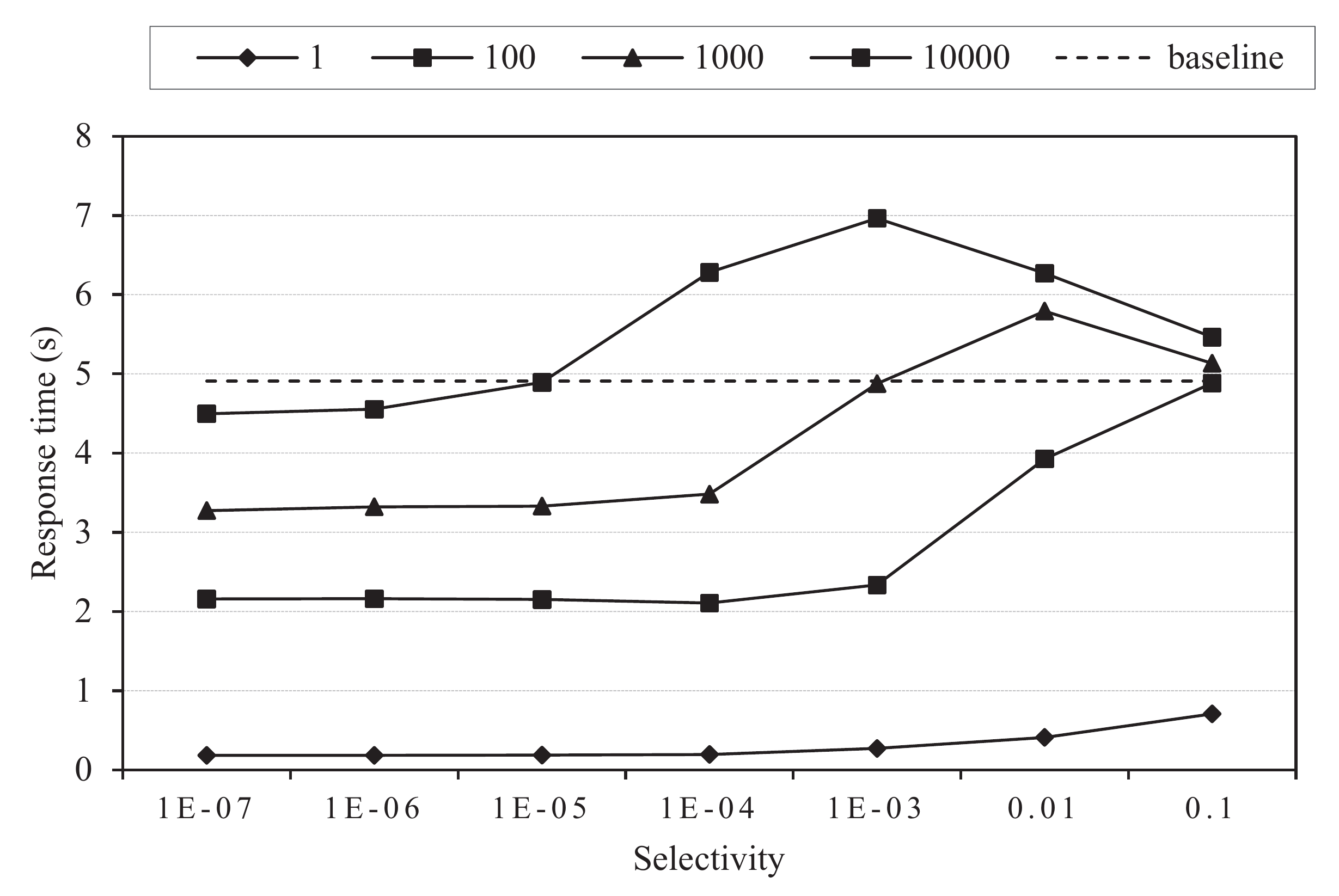}
   \caption{The effect of selectivity on the response time}
   \label{fig:selectivity}
\end{figure}

In order to show the effect of selectivity on the performance of the algorithm, we conducted an experiments with the synthetic dataset and a random query mode. 
The selectivity is manipulated constantly in accordance with the number of the ART query, and the response time of the algorithm for different number of queries is recorded. 
The experimental results are shown in Figure~\ref{fig:selectivity}. 

Firstly, the greater the degree of selection, the greater the initialization time for the first execution of the algorithm will be. 
And the response time increases with the increase of the number of queries. 
Secondly, when the number of queries is constant, the response time increases slowly at first, if the query selectivity is small. 
It increased significantly as the selectivity doubled.
As can be seen from the figure, when the selectivity is 0.001 and 0.01, the response time changes significantly as the number of queries increases.
With the increase of selectivity, the interval between range queries is increased by a large amount compared with the smaller selectivity, and the overhead of ART index initialization and ART range queries for ART cracking algorithm is increased significantly.
Therefore, the efficiency of the ART cracking algorithm is influenced by the selectivity. 
The appropriate selectivity will result in better performance for the ART carcking algorithm, which is beneficial to the construction of the ART index and the query response time.

\subsection{Convergence Speed}\label{subsec:convergence}
As stated in Section~\ref{subsec:art-range}, there are four phases in the ART cracking, \ie initialization, search ART, cracking, and finishing.
Apparently, the overall query efficiency will be greatly improved if the ART index is completely constructed (or converged) as soon as possible.
To measure the convergence degree of the algorithm, the definition of the ART building rate $R$ is given as follows.
\[R = key.size / N * 100\%\]

The tree building rate $R$ is defined to be the ratio of the number of keys indexed in the ART to the amount of data $N$ after certain times of range queries.
The change in the rate of establishment in a unit of time reflects the convergence speed of the algorithm.

The tree building rate $R$ is an important indicator to measure the performance of the algorithm. 
The more proximate the $R$ value is to 1, the closer the generated ART is to the complete index. 
The degree of convergence of the algorithm depends on the query execution. 
On the one hand, whatever the query pattern is, with the increasing number of queries, the algorithm will gradually converge to the complete index. 
Since the hot data will be queried repeatedly, it has a good query performance even though the tree building rate $R$ is low. 
On the other hand, for the queries with higher selectivity, they usually converge faster, \ie the tree building rate increase faster. 
In the extreme case, if the first query selects all tuples, the ART cracking process will degenerate into a standard ART query process, and hence the selectivity will become the primary factor for algorithm convergence.

\begin{figure}[htp!]
   \includegraphics[width=\linewidth]{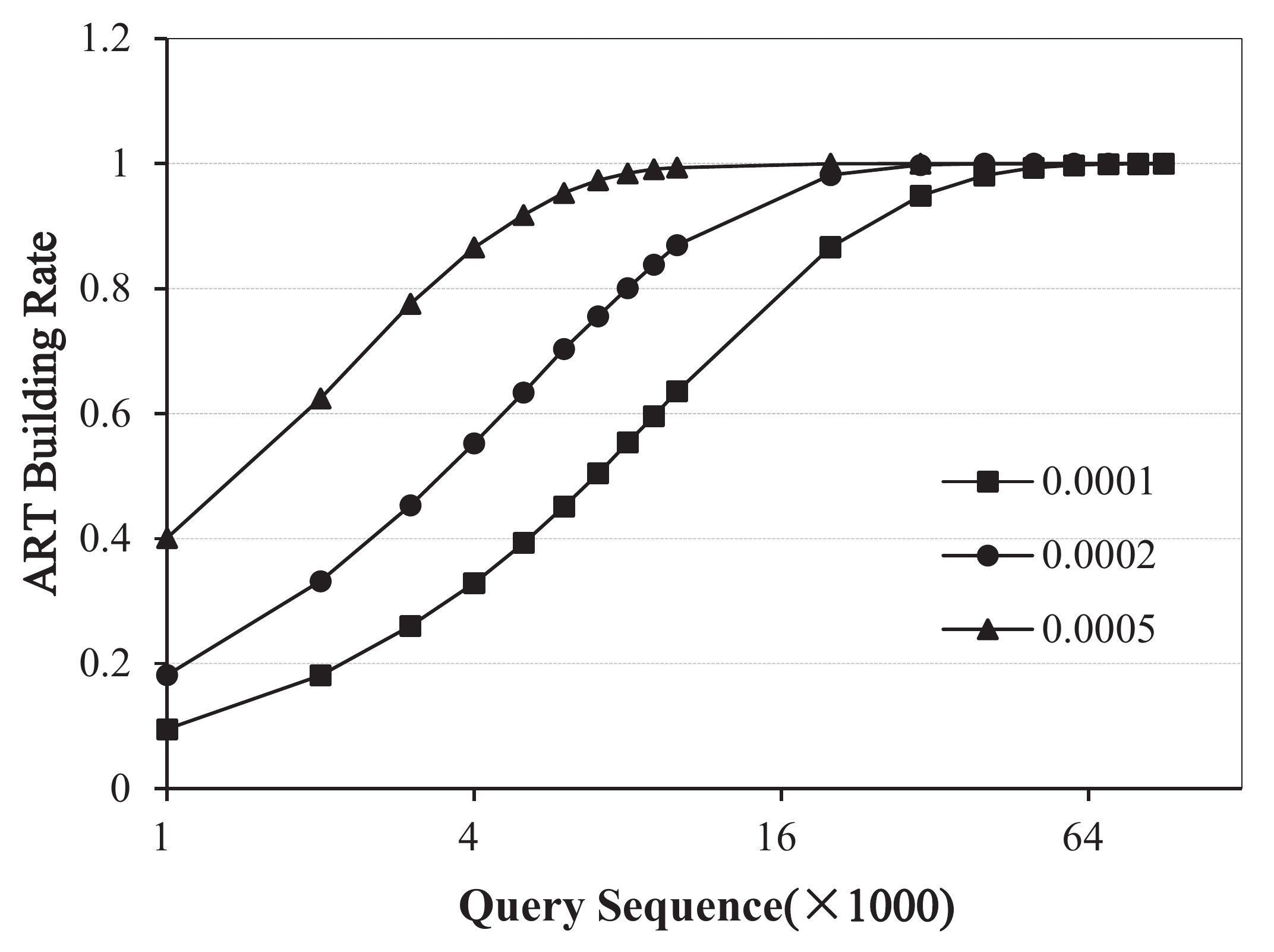}
   \caption{The effect of selectivity on the ART building rate}
   \label{fig:BuildRate}
\end{figure}

In this experiment, we utilized the synthetic dataset with default setting and the random query range mode, and observe the convergence degree of the algorithm under different selection degrees. 
The experimental results are shown in Figure~\ref{fig:BuildRate}. 
We choose three different selectivity. 
When the selectivity is fixed, as the number of queries increases, the tree building rate is gradually approaching 1, and the rate of growth tends to be flat. 
This is mainly because that as the number of queries increases, the overlapping scope of the query range also increases. 
With a fixed selectivity, the number of updated data each time is reduced, so the change rate of the ART building also declines accordingly. 
At the same time, it is noted that as the number of queries increases, the algorithm converges gradually, and the larger the selection rate, the faster the convergence of the algorithm will be. 
When the number of queries reaches 10,000, the selectivity doubles itself and the tree building rate increases by more than 30\%.

\subsection{Range Query Workload Modes}\label{subsec:workload}
The impact of range query workload mode on the standard ART is almost negligible, but it has an important impact on the ART cracking algorithm, which will be shown in this section.
For this experiment, we consider the following range query workloads:
\begin{enumerate}
    \item \emph{Random mode}: the minimum of the range is randomly generated, and the selectivi-ty is fixed.
    \item \emph{Sequential mode}: the minimum of the range increases as the number of queries increases, the maximum value is randomly determined, and the query range may be overlapped.
    \item \emph{Distorted mode}: The top 80\% of the queries are concentrated on 20\% of the data, and the hot data may be queried multiple times.
    \item \emph{Two-way incremental mode}: The query range expands on both the left and right ends on the basis of the previous query.
    \item \emph{One-way incremental mode}: The minimum value of the current query`s range is the maximum value of last query`s range, and the selectivity is fixed.
\end{enumerate}

\begin{figure}[htp!]
   \includegraphics[width=\linewidth]{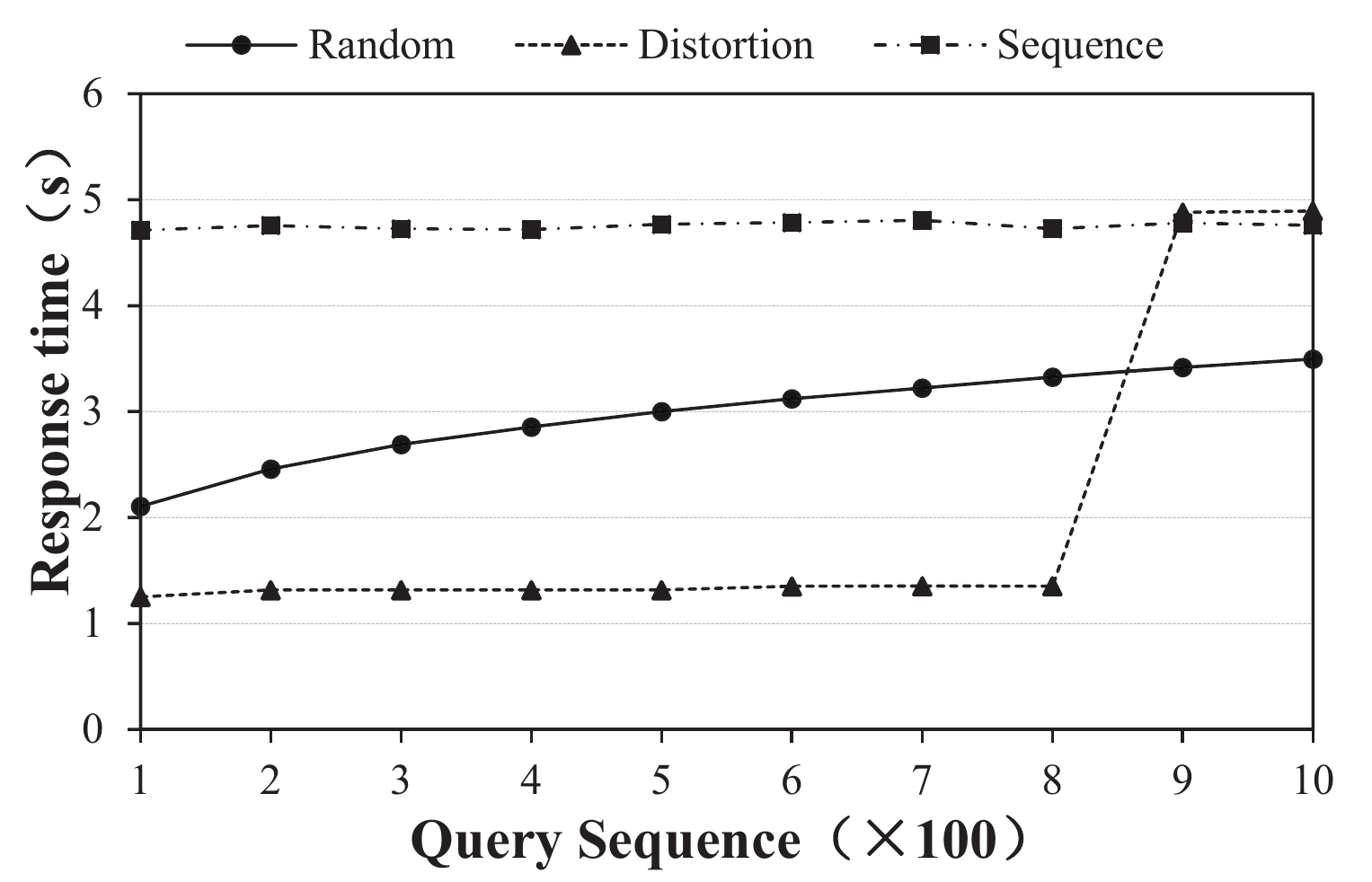}
   \caption{Response time comparisons for different workload modes}
   \label{fig:workload}
\end{figure}

The experimental results are shown in Figure~\ref{fig:workload}. 
For sequential search, the algorithm converges quickly due to the large degree of randomness of the range selection, thus maintaining a high initialization cost. 
For distorted query, the response time is relatively low at the first 80\% of the time, and then increases sharply in the latter part of the query. 
For random queries, as the number of queries increases, the ART cracking algorithm converges slowly and the overall response time is minimal.

\begin{figure}[htp!]
   \includegraphics[scale=0.53]{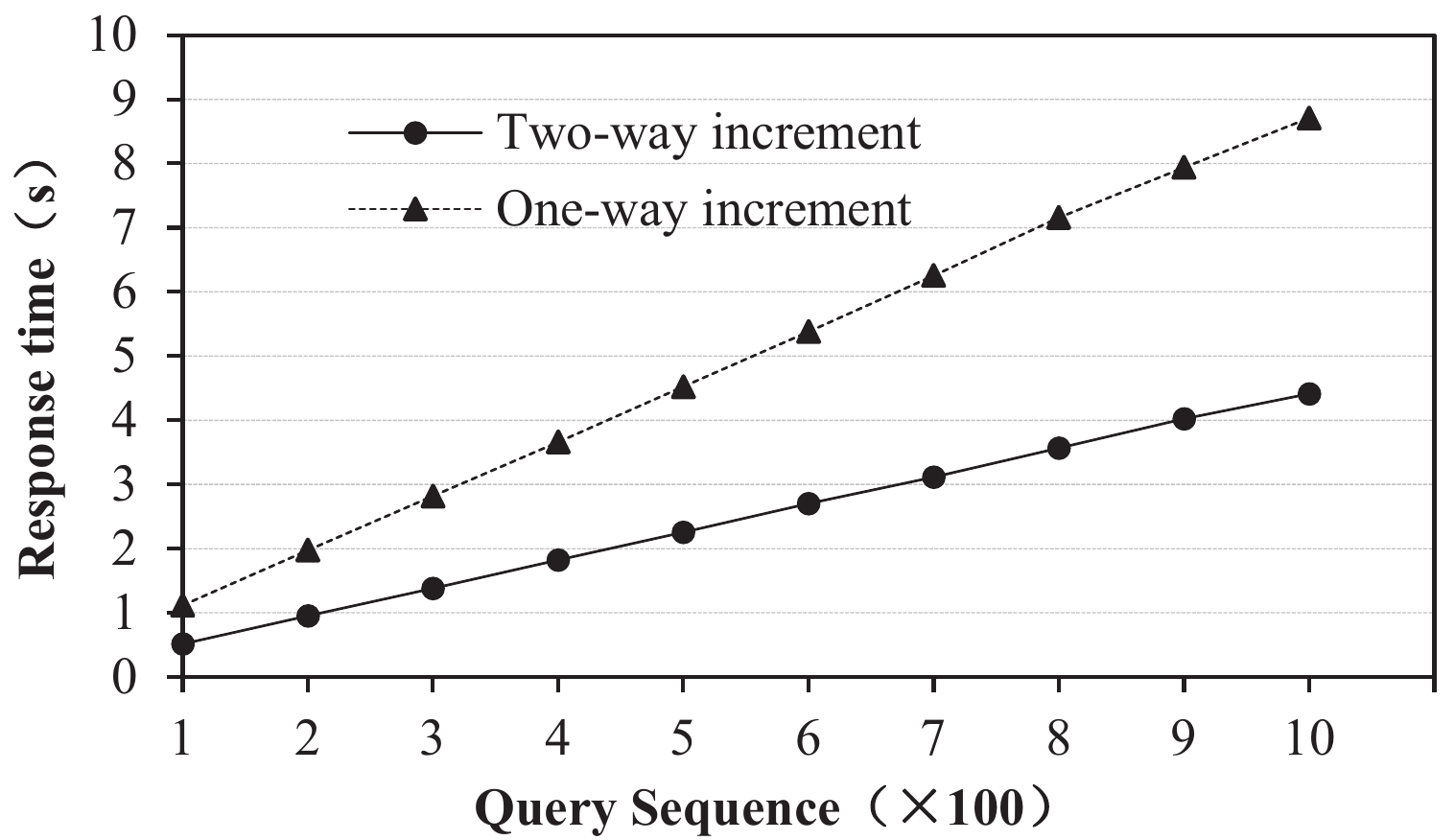}
   \caption{Response time comparisons for two incremental workload modes}
   \label{fig:incremental}
\end{figure}

The experimental results for the two incremental modes are shown in Figure~\ref{fig:incremental}. 
It can be seen that as the range of queries increases, the response time of the two incremental query modes increases almost linearly. 
This is mainly because there is no overlap for the query range of this mode, hence the algorithm degenerates into a standard cracking algorithm which updates the ART each time and never use it in subsequent queries. 
For the two-way incremental mode, the two ends are continuously expanded, and the historical data is stored in the ART, however, each query is inserted at both ends of the range, hence in addition to the algorithm convergence, the overhead of the maintenance algorithm is huge, so the response time increases.

From the two groups of experiments it can be concluded that ART cracking algorithm depends on the change of the workload. 
When the query is in incremental mode, the algorithm has a lower initialization cost, but the subsequent response time is far beyond the ART standard query. 
For other modes, the ART cracking algorithm has both a lower initialization cost and a lower query response time.

\subsection{Per-query response times}\label{subsec:p-query}
\begin{figure}[htp!]
   \includegraphics[scale=0.33]{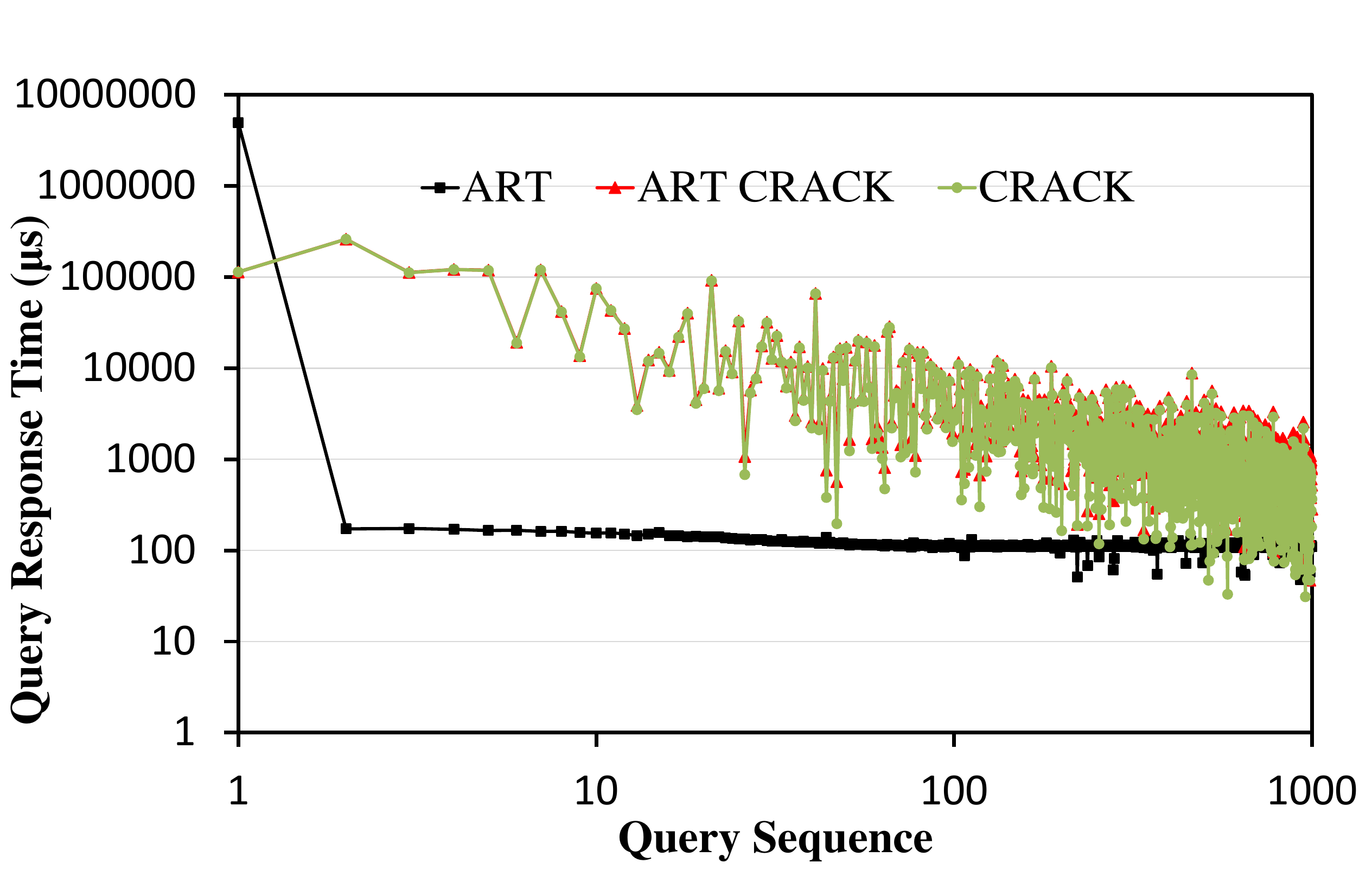}
   \caption{Per-Query Response Time of Different Algorithms}
   \label{fig:pre-response}
\end{figure}

Figure ~\ref{fig:pre-response} shows the per-query response times for different methods in the synthetic dataset and random mode.
Since the available ART implementations do not support bulk loading, the ART cracking algorithm greatly reduces the time required to build full indexes before data is accessed for the first time compared to standard ART.
At that same time, the response time of each query in the ART cracking algorithm is very close to the standard cracking algorithm.

\subsection{Updates}\label{subsec:updates}
\begin{figure}[htp!]
   \includegraphics[width=\linewidth]{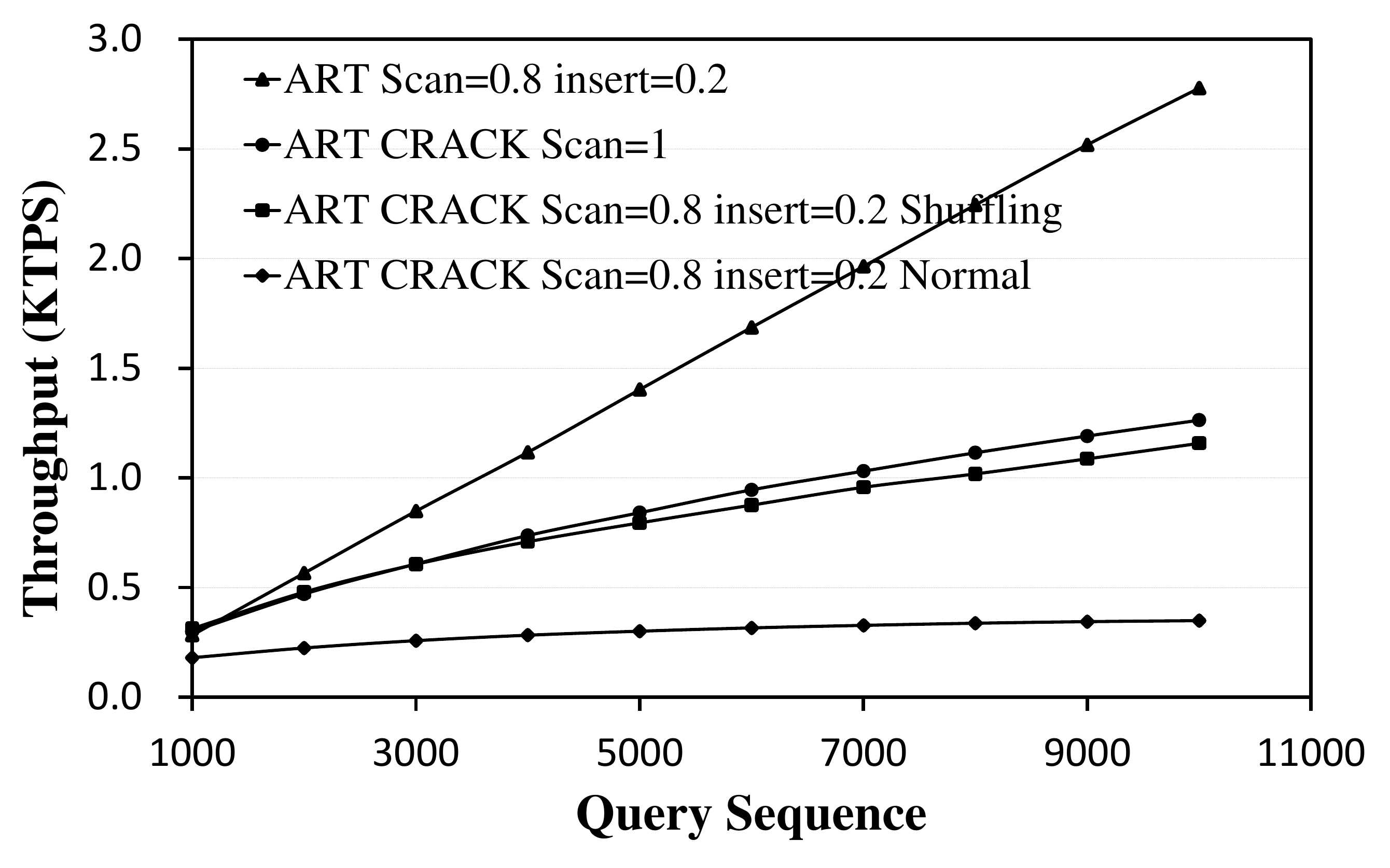}
   \caption{The effect of insertion on two algorithms for the YCSB workload}
   \label{fig:insert-update}
\end{figure}

\begin{figure}[htp!]
   \includegraphics[width=\linewidth]{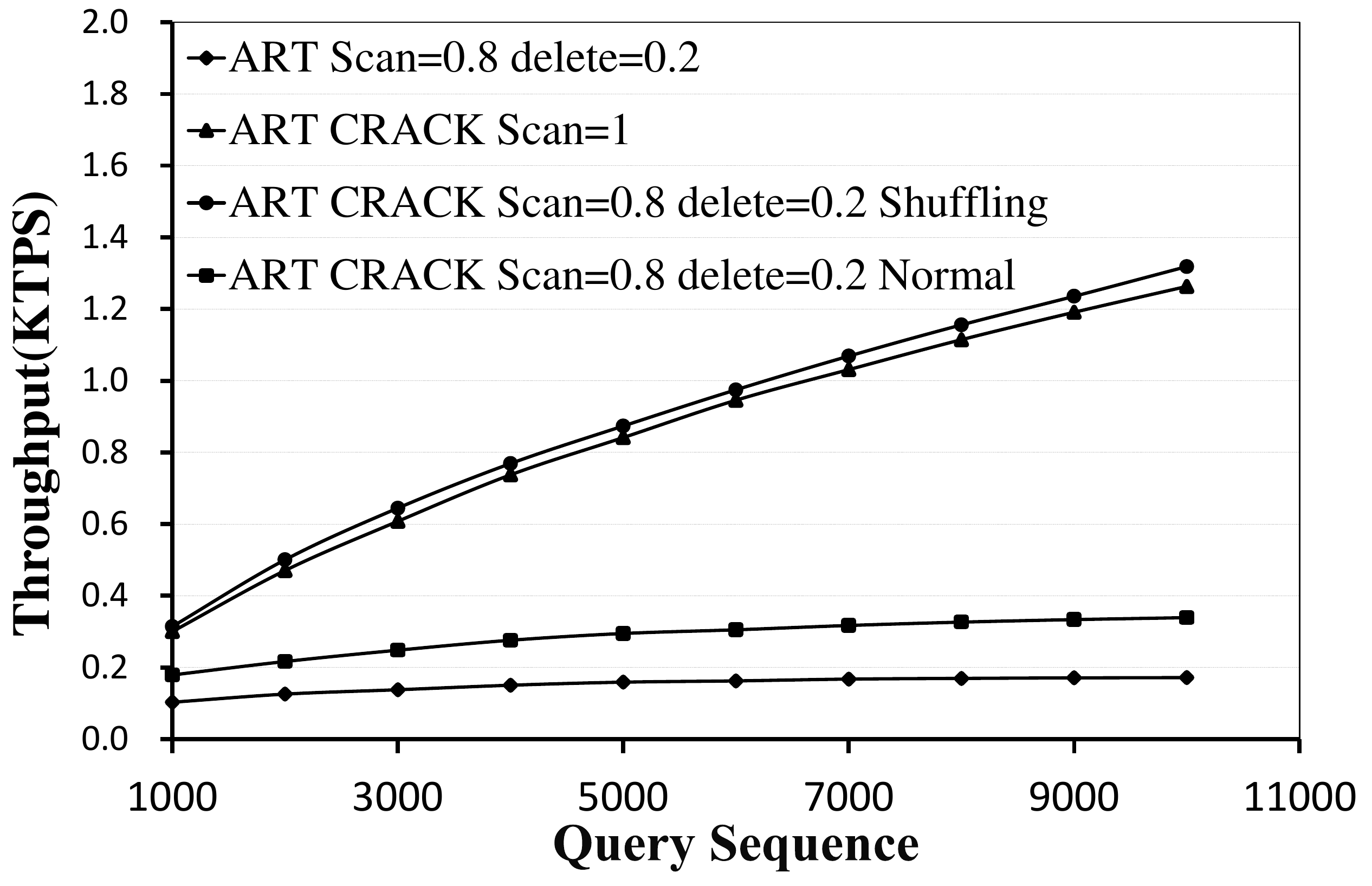}
   \caption{The effect of deletion on two algorithms for the YCSB workload}
   \label{fig:delete-update}
\end{figure}

The above experiments exhibit good performance of the ART cracking algorithm without considering the update.
However, the update of the index in the actual scenario is inevitable.
For this reason, the YCSB benchmark is used to compare the impact of the update on the ART cracking algorithm and the standard ART respectively.
YCSB workload is workload mode e,The experimental results are shown in Figure~\ref{fig:insert-update} and Figure~\ref{fig:delete-update}.

Figure~\ref{fig:insert-update} shows the impact of the insert operation on the standard ART and ART cracking algorithms.
The throughput of the standard ART is significantly higher than that of the ART cracking algorithm because the ART cracking has the overhead of maintaining cracker columns and the range lookup table in addition to maintaining the ART index compared to the standard ART.
At the same time, by using the general insertion method for ART cracking comparison, we can see that using Shuffling method can greatly reduce the impact of insertion on ART cracking algorithm.
The main reason is that the Shuffling method greatly reduces the overhead of inserting cracker columns and the range lookup table maintenance, so that it can achieve almost the same impact as the ART cracking algorithm without insertion.

Figure~\ref{fig:delete-update} shows the impact of the delete operation on the standard ART and ART cracking algorithms. 
The deletion has the most obvious impact on the standard ART, even its throughput is lower than that of the ART cracking with general deletion. 
When the deletion operation is performed, the maximum overhead of both standard ART and ART cracking is to maintain the cracker column.
However, the column slices information is maintained in the ART cracking, the position of the element to be deleted can be quickly located in the cracker column.
Therefore, the throughput of the ART cracking is slightly higher than that of the standard ART. 
Similar to the insertion, using the Shuffling method reduces such overhead.

In summary, by using Shuffing technology, the impact of update on ART cracking algorithm is greatly reduced.

\section{Conclusions and Future work}\label{sec:conclusion}
In order to cope with the huge overhead of in-memory database index creation, this paper proposes an Adaptive Radix Tree (ART) index cracking algorithm based on ART index structure, which improves the instant query response speed by distributing the complete index creation cost to each range query.
The algorithm adapts the conventional database cracking technique to the ART index which enables the query to continuously establish the complete ART index structure during the process to avoid unnecessary initialization overhead and to have a high query efficiency.
The experimental results show the effectiveness of the algorithm.

Undoubtedly, as a case study of the in-memory database index cracking techniques, the experience and lessons learned from the proposed ART index cracking approach are still very preliminary.
Firstly, cracking in-memory index is a simple yet effective technique.
In contrast to building a complete index in the preprocessing stage, cracking the in-memory index is more lightweight.
It does not penalise the first query heavily and also reduces unnecessary initialization time for those cold data seldom visited.
It exhibits better performance when the selectivity is between 0.01\% and 1\% and the workload is random workloads.
Secondly, cracking in-memory index still has the same performance bottlenecks as the standard database cracking, \ie being sensitive to the access patterns when the selectivity is too large or too small, and the workload is too harsh.
This remains a challenge for cracking in-memory index.

The future research will focus on improving the convergence and robustness of cracking in-memory index and the combination of the range lookup table and the ART index to eliminate the access overhead accross different data structure. 
Moreover, a concurrent version of ART cracking is another direction of our efforts. 
Both  \cite{GraefeHIKM2012:concurrent-cracking} and \cite{LeisSKN2016:ART-sync} are good references.

\begin{acks}
Gang Wu is supported by the NSFC (Grant No. 61872072) and the State
Key Laboratory of Computer Software New Technology Open Project Fund (Grant
No. KFKT2018B05).
Baiyou Qiao is supported by the National Key R\&D Program of China (No. 2016YFC1401900).
Donghong Han is supported by the NSFC (Grant No. 61672144).
Guoren Wang is supported by the NSFC (Grant No. U1401256, 61732003, 61332006 and 61729201).
Ye Yuan is supported by the NSFC (Grant No. 61572119 and 61622202) andthe Fundamental Research Funds for the Central Universities (Grant No. N150402005).
\end{acks}

\bibliographystyle{ACM-Reference-Format}
\bibliography{CrackingART-edbt2020}


\begin{thebibliography}{25}


\ifx \showCODEN    \undefined \def \showCODEN     #1{\unskip}     \fi
\ifx \showDOI      \undefined \def \showDOI       #1{#1}\fi
\ifx \showISBNx    \undefined \def \showISBNx     #1{\unskip}     \fi
\ifx \showISBNxiii \undefined \def \showISBNxiii  #1{\unskip}     \fi
\ifx \showISSN     \undefined \def \showISSN      #1{\unskip}     \fi
\ifx \showLCCN     \undefined \def \showLCCN      #1{\unskip}     \fi
\ifx \shownote     \undefined \def \shownote      #1{#1}          \fi
\ifx \showarticletitle \undefined \def \showarticletitle #1{#1}   \fi
\ifx \showURL      \undefined \def \showURL       {\relax}        \fi
\providecommand\bibfield[2]{#2}
\providecommand\bibinfo[2]{#2}
\providecommand\natexlab[1]{#1}
\providecommand\showeprint[2][]{arXiv:#2}

\bibitem[\protect\citeauthoryear{Cooper, Silberstein, Tam, Ramakrishnan, and
  Sears}{Cooper et~al\mbox{.}}{2010}]%
        {CooperSTRS2010:YCSB}
\bibfield{author}{\bibinfo{person}{Brian~F. Cooper}, \bibinfo{person}{Adam
  Silberstein}, \bibinfo{person}{Erwin Tam}, \bibinfo{person}{Raghu
  Ramakrishnan}, {and} \bibinfo{person}{Russell Sears}.}
  \bibinfo{year}{2010}\natexlab{}.
\newblock \showarticletitle{Benchmarking cloud serving systems with {YCSB}}. In
  \bibinfo{booktitle}{\emph{In Proceedings of the 1st {ACM} Symposium on Cloud
  Computing (SoCC 2010)}}. \bibinfo{pages}{143--154}.
\newblock
\urldef\tempurl%
\url{https://doi.org/10.1145/1807128.1807152}
\showDOI{\tempurl}


\bibitem[\protect\citeauthoryear{Graefe, Halim, Idreos, Kuno, and
  Manegold}{Graefe et~al\mbox{.}}{2012}]%
        {GraefeHIKM2012:concurrent-cracking}
\bibfield{author}{\bibinfo{person}{Goetz Graefe}, \bibinfo{person}{Felix
  Halim}, \bibinfo{person}{Stratos Idreos}, \bibinfo{person}{Harumi~A. Kuno},
  {and} \bibinfo{person}{Stefan Manegold}.} \bibinfo{year}{2012}\natexlab{}.
\newblock \showarticletitle{Concurrency Control for Adaptive Indexing}.
\newblock \bibinfo{journal}{\emph{{PVLDB}}} \bibinfo{volume}{5},
  \bibinfo{number}{7} (\bibinfo{year}{2012}), \bibinfo{pages}{656--667}.
\newblock
\urldef\tempurl%
\url{https://doi.org/10.14778/2180912.2180918}
\showDOI{\tempurl}


\bibitem[\protect\citeauthoryear{Graefe and Kuno}{Graefe and Kuno}{2010a}]%
        {Graefe2010:adaptive}
\bibfield{author}{\bibinfo{person}{Goetz Graefe} {and} \bibinfo{person}{Harumi
  Kuno}.} \bibinfo{year}{2010}\natexlab{a}.
\newblock \showarticletitle{Adaptive indexing for relational keys}.
\newblock \bibinfo{journal}{\emph{In Proceedings of the 26rd {IEEE}
  International Conference on Data Engineering (ICDEW2010)}},
  \bibinfo{pages}{69--74}.
\newblock
\urldef\tempurl%
\url{https://doi.org/10.1109/ICDEW.2010.5452743}
\showDOI{\tempurl}


\bibitem[\protect\citeauthoryear{Graefe and Kuno}{Graefe and Kuno}{2010b}]%
        {Graefe2010:self-tuning}
\bibfield{author}{\bibinfo{person}{Goetz Graefe} {and} \bibinfo{person}{Harumi
  Kuno}.} \bibinfo{year}{2010}\natexlab{b}.
\newblock \showarticletitle{Self-selecting, self-tuning, incrementally
  optimized indexes}.
\newblock \bibinfo{journal}{\emph{{EDBT}}}, \bibinfo{pages}{371--381}.
\newblock
\urldef\tempurl%
\url{https://web.eecs.umich.edu/~mozafari/fall2015/eecs584/papers/self-tuning-index.pdf}
\showURL{%
\tempurl}


\bibitem[\protect\citeauthoryear{Halim, Idreos, Karras, and Yap}{Halim
  et~al\mbox{.}}{2012}]%
        {Halim:stochastic}
\bibfield{author}{\bibinfo{person}{Felix Halim}, \bibinfo{person}{Stratos
  Idreos}, \bibinfo{person}{Panagiotis Karras}, {and} \bibinfo{person}{Roland
  H~C Yap}.} \bibinfo{year}{2012}\natexlab{}.
\newblock \showarticletitle{Stochastic database cracking: towards robust
  adaptive indexing in main-memory column-stores}.
\newblock \bibinfo{journal}{\emph{{VLDB} Journal}} \bibinfo{volume}{5},
  \bibinfo{number}{6}, \bibinfo{pages}{502--513}.
\newblock
\urldef\tempurl%
\url{http://vldb.org/pvldb/vol5/p502_felixhalim_vldb2012.pdf}
\showURL{%
\tempurl}


\bibitem[\protect\citeauthoryear{Hankins and Patel}{Hankins and Patel}{2003}]%
        {Hankins2003:csb-tree}
\bibfield{author}{\bibinfo{person}{Richard~A Hankins} {and}
  \bibinfo{person}{Jignesh~M Patel}.} \bibinfo{year}{2003}\natexlab{}.
\newblock \showarticletitle{Effect of node size on the performance of
  cache-conscious B + -trees}.
\newblock \bibinfo{journal}{\emph{measurement and modeling of computer
  systems}} \bibinfo{volume}{31}, \bibinfo{number}{1},
  \bibinfo{pages}{283--294}.
\newblock


\bibitem[\protect\citeauthoryear{Idreos, Groffen, Nes, Manegold, Mullender, and
  Kersten}{Idreos et~al\mbox{.}}{2012}]%
        {IdreosGNMMK2012:MonetDB}
\bibfield{author}{\bibinfo{person}{Stratos Idreos}, \bibinfo{person}{Fabian
  Groffen}, \bibinfo{person}{Niels Nes}, \bibinfo{person}{Stefan Manegold},
  \bibinfo{person}{K.~Sjoerd Mullender}, {and} \bibinfo{person}{Martin~L.
  Kersten}.} \bibinfo{year}{2012}\natexlab{}.
\newblock \showarticletitle{MonetDB: Two Decades of Research in Column-oriented
  Database Architectures}.
\newblock \bibinfo{journal}{\emph{{IEEE} Database Engineering Bulletin}}
  \bibinfo{volume}{35}, \bibinfo{number}{1} (\bibinfo{year}{2012}),
  \bibinfo{pages}{40--45}.
\newblock
\urldef\tempurl%
\url{http://sites.computer.org/debull/A12mar/monetdb.pdf}
\showURL{%
\tempurl}


\bibitem[\protect\citeauthoryear{Idreos, Kersten, and Manegold}{Idreos
  et~al\mbox{.}}{2007a}]%
        {IdreosKM2007:DB-cracking}
\bibfield{author}{\bibinfo{person}{Stratos Idreos}, \bibinfo{person}{Martin~L.
  Kersten}, {and} \bibinfo{person}{Stefan Manegold}.}
  \bibinfo{year}{2007}\natexlab{a}.
\newblock \showarticletitle{Database Cracking}. In \bibinfo{booktitle}{\emph{In
  Proceedings of the Third Biennial Conference on Innovative Data Systems
  Research (CIDR2007)}}. \bibinfo{pages}{68--78}.
\newblock
\urldef\tempurl%
\url{http://cidrdb.org/cidr2007/papers/cidr07p07.pdf}
\showURL{%
\tempurl}


\bibitem[\protect\citeauthoryear{Idreos, Kersten, and Manegold}{Idreos
  et~al\mbox{.}}{2007b}]%
        {IdreosKM2007:updating-cracked}
\bibfield{author}{\bibinfo{person}{Stratos Idreos}, \bibinfo{person}{Martin~L.
  Kersten}, {and} \bibinfo{person}{Stefan Manegold}.}
  \bibinfo{year}{2007}\natexlab{b}.
\newblock \showarticletitle{Updating a cracked database}. In
  \bibinfo{booktitle}{\emph{In Proceedings of the {ACM} {SIGMOD} International
  Conference on Management of Data (SIGMOD2007)}}. \bibinfo{pages}{413--424}.
\newblock
\urldef\tempurl%
\url{https://doi.org/10.1145/1247480.1247527}
\showDOI{\tempurl}


\bibitem[\protect\citeauthoryear{Idreos, Kersten, and Manegold}{Idreos
  et~al\mbox{.}}{2009}]%
        {IdreosKM2009:self-organizing-cs}
\bibfield{author}{\bibinfo{person}{Stratos Idreos}, \bibinfo{person}{Martin~L.
  Kersten}, {and} \bibinfo{person}{Stefan Manegold}.}
  \bibinfo{year}{2009}\natexlab{}.
\newblock \showarticletitle{Self-organizing tuple reconstruction in
  column-stores}. In \bibinfo{booktitle}{\emph{In Proceedings of the {ACM}
  {SIGMOD} International Conference on Management of Data (SIGMOD2009)}}.
  \bibinfo{pages}{297--308}.
\newblock
\urldef\tempurl%
\url{https://doi.org/10.1145/1559845.1559878}
\showDOI{\tempurl}


\bibitem[\protect\citeauthoryear{Idreos, Manegold, Kuno, and Graefe}{Idreos
  et~al\mbox{.}}{2011}]%
        {IdreosMKG2011:merging-cracking}
\bibfield{author}{\bibinfo{person}{Stratos Idreos}, \bibinfo{person}{Stefan
  Manegold}, \bibinfo{person}{Harumi~A. Kuno}, {and} \bibinfo{person}{Goetz
  Graefe}.} \bibinfo{year}{2011}\natexlab{}.
\newblock \showarticletitle{Merging What's Cracked, Cracking What's Merged:
  Adaptive Indexing in Main-Memory Column-Stores}.
\newblock \bibinfo{journal}{\emph{{PVLDB}}} \bibinfo{volume}{4},
  \bibinfo{number}{9} (\bibinfo{year}{2011}), \bibinfo{pages}{585--597}.
\newblock
\urldef\tempurl%
\url{https://doi.org/10.14778/2002938.2002944}
\showDOI{\tempurl}


\bibitem[\protect\citeauthoryear{Kemper and Neumann}{Kemper and
  Neumann}{2011}]%
        {KemperN2011:Hyper}
\bibfield{author}{\bibinfo{person}{Alfons Kemper} {and} \bibinfo{person}{Thomas
  Neumann}.} \bibinfo{year}{2011}\natexlab{}.
\newblock \showarticletitle{HyPer: {A} hybrid OLTP{\&}OLAP main memory database
  system based on virtual memory snapshots}. In \bibinfo{booktitle}{\emph{In
  Proceedings of the 27th International Conference on Data Engineering
  (ICDE2011)}}. \bibinfo{pages}{195--206}.
\newblock
\urldef\tempurl%
\url{https://doi.org/10.1109/ICDE.2011.5767867}
\showDOI{\tempurl}


\bibitem[\protect\citeauthoryear{Kersten and Manegold}{Kersten and
  Manegold}{2005}]%
        {KerstenM2005:cracking-DS}
\bibfield{author}{\bibinfo{person}{Martin~L. Kersten} {and}
  \bibinfo{person}{Stefan Manegold}.} \bibinfo{year}{2005}\natexlab{}.
\newblock \showarticletitle{Cracking the Database Store}. In
  \bibinfo{booktitle}{\emph{In Proceedings of the Second Biennial Conference on
  Innovative Data Systems Research (CIDR2005)}}. \bibinfo{pages}{213--224}.
\newblock
\urldef\tempurl%
\url{http://cidrdb.org/cidr2005/papers/P18.pdf}
\showURL{%
\tempurl}


\bibitem[\protect\citeauthoryear{Kim, Chhugani, Satish, Sedlar, Nguyen,
  Kaldewey, Lee, Brandt, and Dubey}{Kim et~al\mbox{.}}{2010}]%
        {KimCSSNKLBD2010:fast-index}
\bibfield{author}{\bibinfo{person}{Changkyu Kim}, \bibinfo{person}{Jatin
  Chhugani}, \bibinfo{person}{Nadathur Satish}, \bibinfo{person}{Eric Sedlar},
  \bibinfo{person}{Anthony~D. Nguyen}, \bibinfo{person}{Tim Kaldewey},
  \bibinfo{person}{Victor~W. Lee}, \bibinfo{person}{Scott~A. Brandt}, {and}
  \bibinfo{person}{Pradeep Dubey}.} \bibinfo{year}{2010}\natexlab{}.
\newblock \showarticletitle{{FAST:} fast architecture sensitive tree search on
  modern CPUs and GPUs}. In \bibinfo{booktitle}{\emph{In Proceedings of the
  {ACM} {SIGMOD} International Conference on Management of Data (SIGMOD2010)}}.
  \bibinfo{pages}{339--350}.
\newblock
\urldef\tempurl%
\url{https://doi.org/10.1145/1807167.1807206}
\showDOI{\tempurl}


\bibitem[\protect\citeauthoryear{Lehman and Carey}{Lehman and Carey}{1986}]%
        {LehmanC1986:T-tree}
\bibfield{author}{\bibinfo{person}{Tobin~J. Lehman} {and}
  \bibinfo{person}{Michael~J. Carey}.} \bibinfo{year}{1986}\natexlab{}.
\newblock \showarticletitle{A Study of Index Structures for Main Memory
  Database Management Systems}. In \bibinfo{booktitle}{\emph{In Proceedings of
  the Twelfth International Conference on Very Large Data Bases (VLDB1986)}}.
  \bibinfo{pages}{294--303}.
\newblock
\urldef\tempurl%
\url{http://www.vldb.org/conf/1986/P294.PDF}
\showURL{%
\tempurl}


\bibitem[\protect\citeauthoryear{{Leis}, {Kemper}, and {Neumann}}{{Leis}
  et~al\mbox{.}}{2013}]%
        {LeisKN2013:ART-index}
\bibfield{author}{\bibinfo{person}{V. {Leis}}, \bibinfo{person}{A. {Kemper}},
  {and} \bibinfo{person}{T. {Neumann}}.} \bibinfo{year}{2013}\natexlab{}.
\newblock \showarticletitle{The adaptive radix tree: ARTful indexing for
  main-memory databases}. In \bibinfo{booktitle}{\emph{In Proceedings of the
  29th International Conference on Data Engineering (ICDE 2013)}}.
  \bibinfo{pages}{38--49}.
\newblock


\bibitem[\protect\citeauthoryear{Leis, Scheibner, Kemper, and Neumann}{Leis
  et~al\mbox{.}}{2016}]%
        {LeisSKN2016:ART-sync}
\bibfield{author}{\bibinfo{person}{Viktor Leis}, \bibinfo{person}{Florian
  Scheibner}, \bibinfo{person}{Alfons Kemper}, {and} \bibinfo{person}{Thomas
  Neumann}.} \bibinfo{year}{2016}\natexlab{}.
\newblock \showarticletitle{The {ART} of practical synchronization}. In
  \bibinfo{booktitle}{\emph{In Proceedings of the 12th International Workshop
  on Data Management on New Hardware (DaMoN 2016)}}. \bibinfo{pages}{3:1--3:8}.
\newblock
\urldef\tempurl%
\url{https://doi.org/10.1145/2933349.2933352}
\showDOI{\tempurl}


\bibitem[\protect\citeauthoryear{Levandoski, Lomet, and Sengupta}{Levandoski
  et~al\mbox{.}}{2013}]%
        {LevandoskiLS2013a:BwTree}
\bibfield{author}{\bibinfo{person}{Justin~J. Levandoski},
  \bibinfo{person}{David~B. Lomet}, {and} \bibinfo{person}{Sudipta Sengupta}.}
  \bibinfo{year}{2013}\natexlab{}.
\newblock \showarticletitle{The Bw-Tree: {A} B-tree for new hardware
  platforms}. In \bibinfo{booktitle}{\emph{In Proceedings of the 29th {IEEE}
  International Conference on Data Engineering (ICDE2013)}}.
  \bibinfo{pages}{302--313}.
\newblock
\urldef\tempurl%
\url{https://doi.org/10.1109/ICDE.2013.6544834}
\showDOI{\tempurl}


\bibitem[\protect\citeauthoryear{Mao, Kohler, and Morris}{Mao
  et~al\mbox{.}}{2012}]%
        {MaoKM2012:mass-tree}
\bibfield{author}{\bibinfo{person}{Yandong Mao}, \bibinfo{person}{Eddie
  Kohler}, {and} \bibinfo{person}{Robert~Tappan Morris}.}
  \bibinfo{year}{2012}\natexlab{}.
\newblock \showarticletitle{Cache craftiness for fast multicore key-value
  storage}. In \bibinfo{booktitle}{\emph{In Proceedings of the Seventh EuroSys
  Conference 2012 (EuroSys2012)}}. \bibinfo{pages}{183--196}.
\newblock
\urldef\tempurl%
\url{https://doi.org/10.1145/2168836.2168855}
\showDOI{\tempurl}


\bibitem[\protect\citeauthoryear{Rao and Ross}{Rao and Ross}{1999}]%
        {RaoR1999:css-tree}
\bibfield{author}{\bibinfo{person}{Jun Rao} {and} \bibinfo{person}{Kenneth~A.
  Ross}.} \bibinfo{year}{1999}\natexlab{}.
\newblock \showarticletitle{Cache Conscious Indexing for Decision-Support in
  Main Memory}. In \bibinfo{booktitle}{\emph{In Proceedings of 25th
  International Conference on Very Large Data Bases (VLDB1999)}}.
  \bibinfo{pages}{78--89}.
\newblock
\urldef\tempurl%
\url{http://www.vldb.org/conf/1999/P7.pdf}
\showURL{%
\tempurl}


\bibitem[\protect\citeauthoryear{Rao and Ross}{Rao and Ross}{2000}]%
        {Rao2000:csb+-tree}
\bibfield{author}{\bibinfo{person}{Jun Rao} {and} \bibinfo{person}{Kenneth~A.
  Ross}.} \bibinfo{year}{2000}\natexlab{}.
\newblock \showarticletitle{Making B+-Trees Cache Conscious in Main Memory}. In
  \bibinfo{booktitle}{\emph{In Proceedings of the 2000 ACM SIGMOD International
  Conference on Management of Data (SIGMOD2000)}}. \bibinfo{pages}{475--486}.
\newblock


\bibitem[\protect\citeauthoryear{Schuhknecht, Jindal, and Dittrich}{Schuhknecht
  et~al\mbox{.}}{2013}]%
        {FelixMartin2013:uncracked}
\bibfield{author}{\bibinfo{person}{Felix~Martin Schuhknecht},
  \bibinfo{person}{Alekh Jindal}, {and} \bibinfo{person}{Jens Dittrich}.}
  \bibinfo{year}{2013}\natexlab{}.
\newblock \showarticletitle{The Uncracked Pieces in Database Cracking}.
\newblock \bibinfo{journal}{\emph{{VLDB} Journal}} \bibinfo{volume}{7},
  \bibinfo{number}{2}, \bibinfo{pages}{97--108}.
\newblock
\urldef\tempurl%
\url{http://www.vldb.org/pvldb/vol7/p97-schuhknecht.pdf}
\showURL{%
\tempurl}


\bibitem[\protect\citeauthoryear{Schuhknecht, Jindal, and Dittrich}{Schuhknecht
  et~al\mbox{.}}{2016}]%
        {SchuhknechtJD2016:evaluation-cracking}
\bibfield{author}{\bibinfo{person}{Felix~Martin Schuhknecht},
  \bibinfo{person}{Alekh Jindal}, {and} \bibinfo{person}{Jens Dittrich}.}
  \bibinfo{year}{2016}\natexlab{}.
\newblock \showarticletitle{An experimental evaluation and analysis of database
  cracking}.
\newblock \bibinfo{journal}{\emph{{VLDB} Journal}} \bibinfo{volume}{25},
  \bibinfo{number}{1} (\bibinfo{year}{2016}), \bibinfo{pages}{27--52}.
\newblock
\urldef\tempurl%
\url{https://doi.org/10.1007/s00778-015-0397-y}
\showDOI{\tempurl}


\bibitem[\protect\citeauthoryear{Xie, Cai, Chen, Mao, and Zhang}{Xie
  et~al\mbox{.}}{2018}]%
        {XieCCMZ2018:evaluation-imdb-index}
\bibfield{author}{\bibinfo{person}{Zhongle Xie}, \bibinfo{person}{Qingchao
  Cai}, \bibinfo{person}{Gang Chen}, \bibinfo{person}{Rui Mao}, {and}
  \bibinfo{person}{Meihui Zhang}.} \bibinfo{year}{2018}\natexlab{}.
\newblock \showarticletitle{A Comprehensive Performance Evaluation of Modern
  In-Memory Indices}. In \bibinfo{booktitle}{\emph{In Proceedings of the 34th
  {IEEE} International Conference on Data Engineering (ICDE2018)}}.
  \bibinfo{pages}{641--652}.
\newblock
\urldef\tempurl%
\url{https://doi.org/10.1109/ICDE.2018.00064}
\showDOI{\tempurl}


\bibitem[\protect\citeauthoryear{Xie, Cai, Jagadish, Ooi, and Wong}{Xie
  et~al\mbox{.}}{2017}]%
        {XieCJOW2017:PSL}
\bibfield{author}{\bibinfo{person}{Zhongle Xie}, \bibinfo{person}{Qingchao
  Cai}, \bibinfo{person}{H.~V. Jagadish}, \bibinfo{person}{Beng~Chin Ooi},
  {and} \bibinfo{person}{Weng{-}Fai Wong}.} \bibinfo{year}{2017}\natexlab{}.
\newblock \showarticletitle{Parallelizing Skip Lists for In-Memory Multi-Core
  Database Systems}. In \bibinfo{booktitle}{\emph{In Proceedings of the 33rd
  {IEEE} International Conference on Data Engineering (ICDE2017)}}.
  \bibinfo{pages}{119--122}.
\newblock
\urldef\tempurl%
\url{https://doi.org/10.1109/ICDE.2017.54}
\showDOI{\tempurl}


\end{thebibliography}

%
% ...

\end{document}